\documentclass[]{article}

\pdfoutput=1
\usepackage{authblk}
\usepackage{fullpage}
\usepackage{amsmath}

\usepackage{caption}
\usepackage{graphicx}

\usepackage[hidelinks]{hyperref}

\usepackage[round, semicolon]{natbib}

\title{A Procedure for Developing Uncertainty-Consistent Vs Profiles from Inversion of Surface Wave Dispersion Data}
\author[1]{Joseph P. Vantassel}
\author[1]{Brady R. Cox}
\affil[1]{The University of Texas at Austin}

\begin{document}

\maketitle

\begin{abstract}
Non-invasive surface wave methods have become a popular alternative to traditional invasive forms of site-characterization for inferring a site's subsurface shear wave velocity (Vs) structure. The advantage of surface wave methods over traditional forms of site characterization is that measurements made solely at the ground surface can be used routinely and economically to infer the subsurface structure of a site to depths of engineering interest (20--50 m), and much greater depths (\textgreater 1 km) in some special cases. However, the quantification and propagation of uncertainties from surface wave measurements into the Vs profiles used in subsequent engineering analyses remains challenging. While this has been the focus of much work in recent years, and while considerable progress has been made, no approach for doing so has been widely accepted, leading analysts to either address the propagation of uncertainties in their own specialized manner or, worse, to ignore these uncertainties entirely. In response, this paper presents an easy-to-implement, effective, and verifiable method for developing uncertainty-consistent Vs profiles from inversion of surface wave dispersion data. We begin by examining four approaches presented in the literature for developing suites of Vs profiles meant to account for uncertainty present in the measured dispersion data. These methods are shown to be deficient in three specific ways. First, all approaches are shown to be highly sensitive to their many user-defined inversion input parameters, making it difficult/impossible for them to be performed repeatedly by different analysts. Second, the suites of inverted Vs profiles, when viewed in terms of their implied theoretical dispersion data, are shown to significantly underestimate the uncertainty present in the experimental dispersion data, though some may appear satisfactory when viewed purely qualitatively. Third, if the uncertainties in the implied theoretical dispersion data were to be examined quantitatively, which has not been done previously, there is no obvious remedy available to the analyst to resolve any inconsistency between the measured and inverted dispersion uncertainty. Therefore, a new approach is proposed that seeks to remedy these shortcomings. First, beyond appropriate considerations that must be given to all inversions, the method is governed by only one user-defined input parameter, to which it is not overly sensitive. Second, it is shown to produce suites of Vs profiles whose theoretical dispersion data quantitatively reproduce the uncertainties in the experimental dispersion data. Third, the final step of the procedure requires the analyst to compare the measured and inverted dispersion uncertainties quantitatively, and should the analyst find the results to be lacking, clear guidance is provided on the actions necessary to produce Vs profiles whose theoretical dispersion data better account for the experimental uncertainty. Using two synthetic tests and a real-world example, the procedure is shown to produce suites of Vs profiles that accurately capture the site's Vs structure, while rigorously propagating the dispersion data's uncertainty through the inversion process.
\end{abstract}

\pagebreak

\section{Introduction}
Surface wave methods have become a popular alternative to traditional invasive forms of site characterization for measuring a site's subsurface shear wave velocity (Vs) structure.  Vs, which is directly related to the site's small-strain shear stiffness ($G_{max}=\rho Vs^2$, where $\rho$ refers to mass density), is a critical parameter in many seismic hazard analyses, such as ground response analyses \citep{foti_non-uniqueness_2009, rathje_influence_2010, teague_site_2016, passeri_influence_2019} and liquefaction-triggering \citep{andrus_liquefaction_2000, kayen_shear-wave_2013, wood_vs-based_2017}. The performance of surface wave testing is traditionally broken into three stages: acquisition, processing, and inversion \citep{foti_surface_2015}. The acquisition stage involves non-invasively (i.e., from the ground surface) measuring surface waves as they propagate through a site. This can be accomplished using active-source methods, where surface waves are generated by the experimenters, and/or passive-wavefield methods, where sensors are left undisturbed to record ambient surface waves. The processing stage transforms these time-domain recordings to measurements of the site's dispersive properties, which in this context describes how the site's surface wave phase velocity changes as a function of frequency (or equivalently wavelength). Importantly, this measurement of the site's dispersive properties should include site-specific estimates of frequency-dependent uncertainty, discussed in detail later. For clarity of expression, we will refer to the measurement of the site's dispersive properties from the processing stage as the experimental dispersion data. Importantly, when surface wave processing is performed by an experienced analyst, the experimental dispersion data is robustly determined, with uncertainties that typically range from 5\% -- 10\% coefficient of variation (COV) between analysts \citep{cox_synthesis_2014, garofalo_interpacific_2016}. This consistency led Griffiths et al. (\citeyear{griffiths_mapping_2016, griffiths_surface-wave_2016}) and Teague et al. (\citeyear{teague_development_2018}) to aptly refer to the experimental dispersion data as part of a site's ``signature''. The final stage of surface wave testing is inversion. In this stage, numerical search algorithms are used to identify one-dimensional (1D) layered earth models whose theoretical dispersion curves, determined through an analytical forward problem, best fit the experimental dispersion data. It is important for the reader to understand that, while throughout this paper we focus primarily on Vs, as it has the greatest impact on the theoretical dispersion curve \citep{wathelet_array_2005} and is of primary importance to subsequent engineering analyses, the computation of a theoretical dispersion curve (and therefore surface wave inversion) requires the definition of an entire 1D ground model. These ground models are composed of a stack of layers described by their thickness (H), Vs, compression-wave velocity (Vp), and $\rho$. The models whose theoretical dispersion curves best fit the experimental dispersion data, as determined by a misfit function, are considered to be the most likely representations of the site's subsurface conditions.

As the primary focus of this paper is on the consistent propagation of experimental dispersion data uncertainty into the Vs profiles resulting from inversion, it is important to first discuss the types and potential sources of these uncertainties. Uncertainties can be broadly grouped into two categories commonly used in probabilistic seismic hazard analysis (PSHA): epistemic and aleatory. Epistemic uncertainty describes those unknowns which stem from a lack of knowledge. In surface wave testing, sources of epistemic uncertainty include, among others, the selected dispersion processing wavefield transformation and the number of 1D subsurface layers used during inversion. Aleatory uncertainty, or sometimes referred to as aleatory variability, is the result of inherit randomness within the quantity being measured. When discussing site characterization in particular, aleatory uncertainty is classically used to define how the material properties in the subsurface change in three-dimensional space (i.e., randomness with space, or spatial variability). Of course, this could also be considered epistemic uncertainty because it stems from a lack of knowledge rather than from any inherit randomness within the subsurface (i.e., randomness with time). Being able to consider subsurface variability as either epistemic or aleatory in nature hints that, for most real-world applications, the two uncertainty ``categories'' are not as distinct as their names may indicate. Regardless, it is to be expected that all experimental dispersion data contains some amount of uncertainty and that this uncertainty is in part epistemic (dependent largely on the quantity and quality of the data acquired) and part aleatory (dependent on the complexity of the subsurface and the spatial extents of the surface wave arrays). And, while it is difficult-to-impossible to separate epistemic and aleatory uncertainties in the experimental dispersion data, it is of paramount concern that the surface wave analyst attempt to quantify the combined uncertainty on a site-by-site basis (discussed briefly next) and then propagate that uncertainty through the inversion process in order to develop uncertainty-consistent Vs profiles. 

\begin{figure}[t]
	\includegraphics[width=\textwidth]{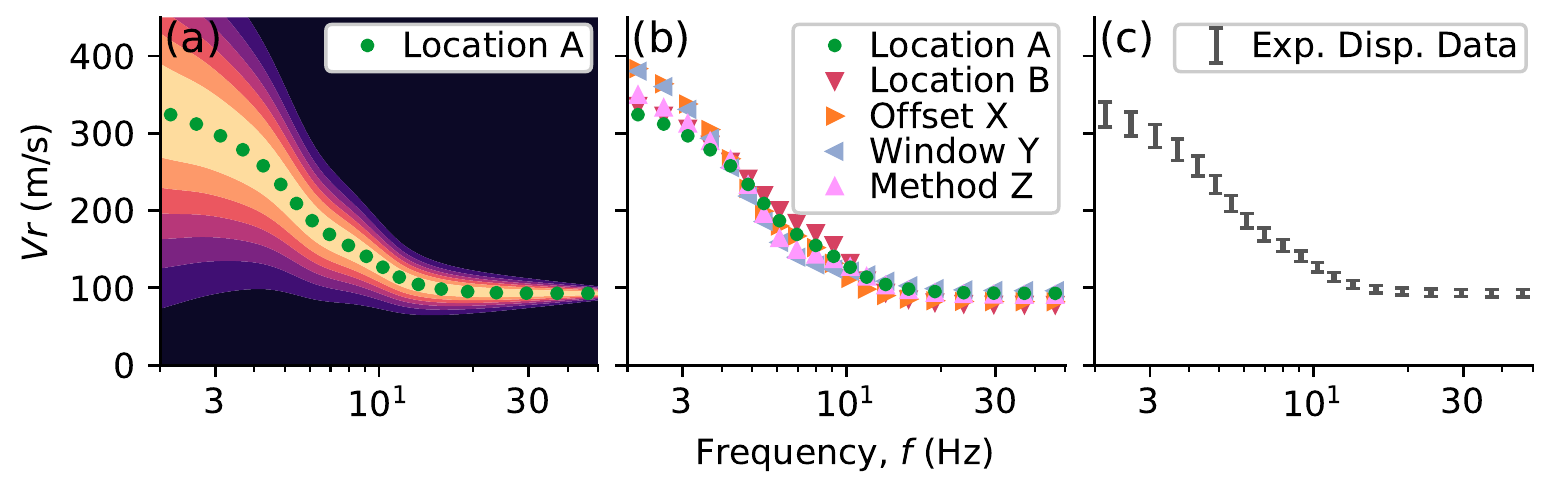}
	\caption{Schematic illustrating one possible procedure for developing experimental dispersion data with measures of uncertainty in terms of the site's Rayleigh wave velocity (Vr). This involves: (a) the processing of recorded waveforms from a single location, experimental array setup, source offset/noise time window, and wavefield transformation method, (b) the modification and repetition of the procedure illustrated in (a) to produce estimates of the site's dispersion uncertainty, and (c) the synthesis of these estimates of uncertainty into a statistical representation called the site's experimental dispersion data.}
	\label{fig:1}
\end{figure}

As this study begins under the assumption that site-specific estimates of dispersion uncertainty have been established, how this might be accomplished deserves a brief discussion. However, before doing so, it is important to acknowledge that while many previous researchers have developed procedures for accounting for uncertainty in experimental data \citep{lai_propagation_2005, foti_non-uniqueness_2009,  cox_surface_2011, teague_development_2018} no single procedure has been accepted widely into practice, and therefore the example presented here is but one of many potential alternatives that may be used. To facilitate a more practical discussion, the example presented here is shown in Figure \ref{fig:1}. Figure \ref{fig:1}a illustrates a measurement of a site's dispersive properties at some location termed ``Location A''. This measurement is directly linked to a given location (i.e., Location A), experimental setup (e.g., array configuration), wavefield recording (e.g., source offset or noise time window), and wavefield processing method (e.g., frequency-wavenumber transformation). The light-colored portions of Figure \ref{fig:1}a show high surface wave power and the dark-colored portions low surface wave power. The maximum power at each frequency is selected as the representative experimental dispersion data for that location, setup, wavefield recording, and processing method. To then develop meaningful statistics that incorporate the uncertainties previously discussed, the procedure shown in Figure \ref{fig:1}a should be modified and/or repeated in a systematic manner to encompass reasonable combinations of these contributing sources. As shown schematically in Figure \ref{fig:1}b, these combinations may include performing the test at various locations across a site (i.e., Location B), using various source locations for active-source experiments (i.e., Offset X), various time-windows for passive-wavefield experiments (i.e., Window Y), and multiple wavefield processing methods (i.e., Method Z). Finally, once these sources of uncertainty have been accounted for in the form of experimental dispersion data (i.e., Figure \ref{fig:1}b) they may be summarized into a statistical representation at each frequency (or wavelength), as shown in Figure \ref{fig:1}c. A previous study by Lai et al. (\citeyear{lai_propagation_2005}) showed that the uncertainty in experimental dispersion data for active-source experiments based on multiple source impacts was normally distributed, therefore, it is common to represent the Rayleigh wave velocity (Vr) at each frequency with a mean and standard deviation. However, it is important to note that at some sites with significant lateral variability, such as the Garner Valley site examined by Teague et al. (\citeyear{teague_measured_2018}), it may be more appropriate to develop alternative dispersion data sets (one per location) with their own experimental uncertainty and invert them separately rather than trying to represent all of the dispersion data with a single statistical distribution. As a more in depth discussion of developing experimental dispersion data is beyond the scope of this paper, the reader is encouraged to carefully consider the impact of how defining their experimental dispersion uncertainty will affect the resulting distribution of Vs profiles and whether those profiles will correctly communicate the measured uncertainty.

While not the primary focus of this work, it is important to briefly mention some of the methods currently being used to account for Vs uncertainty in seismic hazard studies and highlight why developing uncertainty-consistent Vs profiles from experimental dispersion data is of such critical importance. At present, most important seismic design projects utilize some form of Vs randomization to account for Vs uncertainty in ground response analyses, where a baseline Vs profile is randomized (i.e., manipulated) using a statistical procedure. The most commonly used of these randomization approaches is that proposed by Toro (\citeyear{toro_probabilistic_1995}), leading to its adoption in the guidelines for hazard-consistent one-dimensional ground response analyses \citep{stewart_guidelines_2014} and the design of critical facilities \citep{epri_seismic_2012}. However, multiple studies \citep{griffiths_mapping_2016, griffiths_surface-wave_2016, teague_measured_2018, teague_site_2016} have expressed serious concern that the blind application of this type of Vs randomization can result in highly unrealistic ground models that do not fit the experimental site signature and an inability of such models to accurately predict ground response. The reader will note that a new Vs randomization approach has recently been proposed by Passeri et al. (\citeyear{passeri_new_2020}). However, the authors have not yet evaluated it rigorously to determine if it produces Vs profiles that are more consistent with the site signature. Importantly, this paper, while not discussing Vs randomization directly, offers an alternative to Vs randomization by propagating measured, site-specific uncertainty into suites of Vs profiles obtained from surface wave testing. These suites of profiles can then be used either directly to address Vs uncertainty in subsequent engineering analyses (e.g., ground response analyses) or, if the use of Vs randomization is still desired, to better inform its many unknown input parameters with site-specific and uncertainty-consistent values. While this study will focus primarily on the development of the former, it is important that the reader be aware of the wider implications of the proposed method on greater engineering practice.

This study begins by examining common approaches found in the literature for developing uncertainty-consistent Vs profiles from the inversion of experimental surface wave dispersion data. Using a synthetic dataset, these approaches from the literature are shown to yield suites of subsurface models whose theoretical dispersion curves severely underestimate the experimental dispersion data's uncertainty, resulting in Vs profile which underestimate the site's uncertainty. This is then followed by the presentation of a new procedure for rigorously propagating measured experimental dispersion uncertainty through the inversion process to obtain suites of Vs profiles that more accurately represent Vs uncertainty at the site. This new procedure is applied to the same synthetic dataset as the literature-based approaches and is shown to be quantitatively superior, and able to precisely propagate the experimental dispersion uncertainty into the resulting Vs profiles. The new procedure is then extended to incorporate epistemic uncertainty in the inversion's layering parameterization using the same synthetic example, and again shows excellent results. The study concludes with the application of the method at a real site where it is shown to produce suites of Vs profiles which agree favorably with a borehole Vs profile while simultaneously capturing the experimental dispersion data's uncertainty.

\section{Common Approaches used to Account for Vs Uncertainty in Surface Wave Inversion}

To illustrate the problem this paper proposes to solve, we first examine several approaches from the literature that have been used to account for Vs uncertainty in surface wave inversion. All of the approaches mentioned below have a basic commonality, in that they first search through large numbers (often tens-of-thousands to millions) of trial layered-earth models to find a significant number of models with an acceptable fit to the experimental dispersion data. The acceptability of a model is typically judged using a misfit function that quantifies the goodness-of-fit between the theoretical dispersion curve for a given model and the experimental dispersion data. Due to the non-uniqueness of the inverse problem and the experimental dispersion data's uncertainty, it is often possible to find thousands- to tens-of-thousands of acceptable models that range from subtly to significantly different. From these acceptable models, a subset are selected to account for uncertainty in Vs. The discussion below will focus primarily on how different studies obtained suites of acceptable models, and how they then selected from those to estimate Vs uncertainty.

Before presenting the various approaches that have been used to account for Vs uncertainty in surface wave inversion, it is important to briefly discuss the details surrounding surface wave inversion, as these details will be important for understanding the results presented later. First, all of the approaches presented in the literature utilize large numbers of trial models from a global-search inversion algorithm to find suites of acceptable models. Global-search algorithms vary in their implementation and may search for acceptable models in several different ways, including randomly (i.e., pure Monte-Carlo), with the aid of some optimization algorithm (i.e., pure optimization), or by using a combination of the two. By far, the most popular tool for performing global-search surface wave inversion is the Dinver module \citep{wathelet_surface-wave_2004} of the open-source software Geopsy \citep{wathelet_geopsy_2020}. As a testament to its popularity, Dinver was the inversion algorithm of choice in all but one of the works discussed below. Of particular relevance to the discussion below, and worth discussing here, is the misfit function proposed by Wathelet et al. (\citeyear{wathelet_surface-wave_2004}) and implemented in Dinver. The Wathelet et al. (\citeyear{wathelet_surface-wave_2004}) misfit function can be described as a root-mean-square error normalized by the experimental dispersion uncertainty \citep{yust_epistemic_2018}. This gives the misfit function a useful physical interpretation, as it represents on average, across all frequencies/wavelengths, how far (in number of standard deviations) a theoretical dispersion curve strays from the experimental dispersion data. For example, a theoretical dispersion curve with a misfit of 1.0 can be understood as a curve that on average is one standard deviation away from the mean. With these details in mind, it is now possible to discuss the approaches presented in the literature to account for Vs uncertainty in surface wave inversion.

To develop suites of acceptable layered-earth models, Wathelet et al. (\citeyear{wathelet_surface-wave_2004}) and Wathelet (\citeyear{wathelet_improved_2008}) performed pure-optimization inversions considering many (over 100,000) trial models. For convenience of expression, and in order to be consistent with common vernacular, we will refer to a pure optimization inversion as a minimum misfit of zero (M0) inversion. M0 inversions are named as such because the goal of the inversion algorithm is to find a model with a misfit equal to zero (i.e., one whose theoretical dispersion curve perfectly matches the mean experimental dispersion data). In the studies noted above, from the many trial models searched, all models with a dispersion misfit value less than 1.0 (tens of thousands in most cases) were selected as a means to propagate the experimental dispersion data's uncertainty into the resulting Vs profiles. This approach has three shortcomings: (1) it may result in a highly variable number of acceptable profiles, depending on the number of trial models attempted in the inversion and the subjective user-defined quality threshold (i.e., misfit less than 1.0 in this case),  (2) for inversions with many models below the threshold (e.g., tens-of-thousands) it may readily become computationally unmanageable to again propagate the Vs uncertainty into subsequent engineering analyses (e.g., ground response analyses), and (3) the use of an M0 inversion will likely cause the majority of the theoretical dispersion curves to be clustered around the mean of the experimental dispersion data (i.e., a misfit of zero) rather than following the distribution of the experimental dispersion data uncertainty.

Foti et al. (\citeyear{foti_non-uniqueness_2009}) used a pure Monte-Carlo global-inversion algorithm (i.e., not Dinver) to develop suites of acceptable models (more than 50,000 in the examples presented) whose theoretical dispersion curves fit the experimental dispersion data. To select a more manageable subset of models, they sampled from these large suites using a statistical test to select models which could be considered equivalent in term of their fit to the experimental dispersion data, given its uncertainties. While the selection process reduced the computational burden of having to consider ten-of-thousands of Vs profiles, it still maintained the disadvantage of the previous technique that the number of profiles varied between application, ranging between 6 and 270 Vs profiles for the cases presented. Furthermore, the theoretical dispersion curves from the selected models can be seen to underestimate the experimental dispersion data uncertainty, even for the most favorable case where 270 profiles were selected by the statistical test.

Hollender et al. (\citeyear{hollender_characterization_2018}), when developing statistics for the time-averaged shear wave velocity in the upper 30 m (Vs30), used an ``acceptable-misfit'' approach [i.e., an inversion where the minimum misfit was not zero (M0+)] to develop a large suite of acceptable models (over 50,000) to represent the experimental dispersion data uncertainty. From those models, they randomly extracted at least 6000 Vs profiles to develop statistics on Vs30. They then selected a smaller set of 33 representative Vs profiles, whose uncertainty in Vs30 matched that of the randomly selected 6000 profiles, for use in subsequent analyses. While this approach ensured an equal and manageable number of profiles (i.e., 33) for each site, and permitted the propagation of the estimated uncertainty in Vs30, it does not guarantee that the initial large suite of acceptable models properly accounted for the experimental dispersion data's uncertainty. Thereby, potentially allowing the subsequent uncertainties (i.e., the Vs30 uncertainty) upon which they are based to not fully capture the site-specific uncertainty in the experimental dispersion data.

This final section groups a number of studies that are similar, in that they all use a M0 inversion to develop their acceptable suites of models from which to select some fixed number of lowest misfit models to represent Vs uncertainty. Di Giulio et al. (\citeyear{garofalo_interpacific_2016}) selected the 100 lowest misfit profiles out of over a million trial models. Griffiths et al. (\citeyear{griffiths_mapping_2016, griffiths_surface-wave_2016}) and Teague and Cox (\citeyear{teague_site_2016}) chose 50 randomly selected models from the 1000 lowest misfit models obtained from hundreds-of-thousands of trial models, as the 50 randomly selected models were shown to have similar statistical properties as the 1000 best, but were more computationally manageable. Cox and Teague (\citeyear{cox_layering_2016}), Teague et al. (\citeyear{teague_development_2018}) and Deschenes et al. (\citeyear{deschenes_development_2018}) presented the 1000 lowest misfit Vs profiles from hundreds-of-thousands of trial models. Vantassel et al. (\citeyear{vantassel_mapping_2018}), Cox and Vantassel (\citeyear{cox_dynamic_2018}), and Yust et al. (\citeyear{yust_epistemic_2018}) presented the 100 lowest misfit Vs profiles from hundreds-of-thousands of trial models considered to account for Vs uncertainty, as these 100 lowest misfit Vs profiles were shown to be statistically similar to the 1000 lowest misfit Vs profiles, but were more manageable. This approach (i.e., selecting some number of the lowest misfit/``best'' models) resolves the issues of variability between analyses, avoids potentially large numbers of profiles, and is relatively simple. However, this approach is still deficient in two specific ways: (1) it relies on an M0 inversion which, as mentioned previously, will likely produce profiles with theoretical dispersion that is clustered around the experimental dispersion data's mean rather than being properly distributed, and (2) the uncertainty accounted for in the 100 or 1000 best models is indirectly tied to the number of trial models attempted by the analyst (i.e., different apparent uncertainty will results whether the analyst uses 1000, 10,000 or 1,000,000 trial models).

In summary, all of the presented approaches follow a basic two-step process. First, a large number of acceptable models are obtained using a global-search algorithm, and second, some subset of profiles are selected from the acceptable models as a basis for assessing Vs uncertainty. In all of the presented approaches we observe a dependence on two basic assumptions: (1) that the large suite of acceptable models (i.e., from the first step) properly accounts for the experimental dispersion data's uncertainty, and (2) that the selection process will guarantee a set of Vs profiles which rigorously propagates the uncertainty into subsequent analyses. The remainder of this section is devoted to quantitatively assessing the veracity of these assumptions by examining the effectiveness of these approaches.

To quantitatively assess the ability of the approaches from the literature to propagate experimental dispersion data uncertainty into the resulting Vs profiles, four variations were applied to an experimental dispersion dataset. The experimental dispersion data is taken from a large synthetic study focused on the performance of surface wave inversion \citep{vantassel_swinvert_2021}. The data itself has been published as one of twelve surface wave inversion benchmarks, which is publically available on the DesignSafe-CI \citep{vantassel_surface_2020}. While a full detailed discussion is provided in the previous references, in short, the synthetic data was developed by taking the theoretical dispersion curve from an assumed ground model, resampling it in log-wavelength, and assuming a normal distribution in Rayleigh wave velocity (Vr) with a coefficient of variation (COV) of 0.05. A COV of 0.05 was based on typical experimental dispersion data uncertainty values from several blind analyst studies \citep{cox_synthesis_2014, garofalo_interpacific_2016}. For reference, the experimental dispersion data is the same as that shown in Figure \ref{fig:1}c. The proposed methods to be considered were selected to address two primary questions of interest, namely: (1) what effect, if any, does the type of inversion (pure optimization, pure random, or some combination) have on the resulting acceptable models, and (2) how should the analyst sample from the acceptable models. With regard to the type of inversion preformed, we consider two inversion alternatives: (a) a pure-optimization (i.e., M0) inversion, as this approach is the most commonly used in the literature, and (b) a combined approach that is partly optimized and partly random. This second approach will be referred to as a minimum misfit of 1.0 (M1) inversion because the inversion algorithm is forced to search randomly in those regions of the model space where the calculated misfit is below 1.0. A pure-random search was not explicitly considered here, as it is expected to produce similar results to the M1 inversion at greater computational expense. With regard to how the representative Vs profiles should be sampled from the acceptable inversion models, we consider two alternatives: (a) the 100 lowest misfit/``best'' models (b100), as this was the most popular approach in the literature, and (b) 100 randomly selected models from all models with a misfit less than 1.0 (n100). Note that the n100 alternative was a necessary adaptation to the approach of selecting all profiles below the misfit threshold of 1.0 to ensure a computationally manageable number of profiles and a fair comparison to the b100 profiles.  

\begin{figure}[!t]
	\centering
	\includegraphics[width=5.5in]{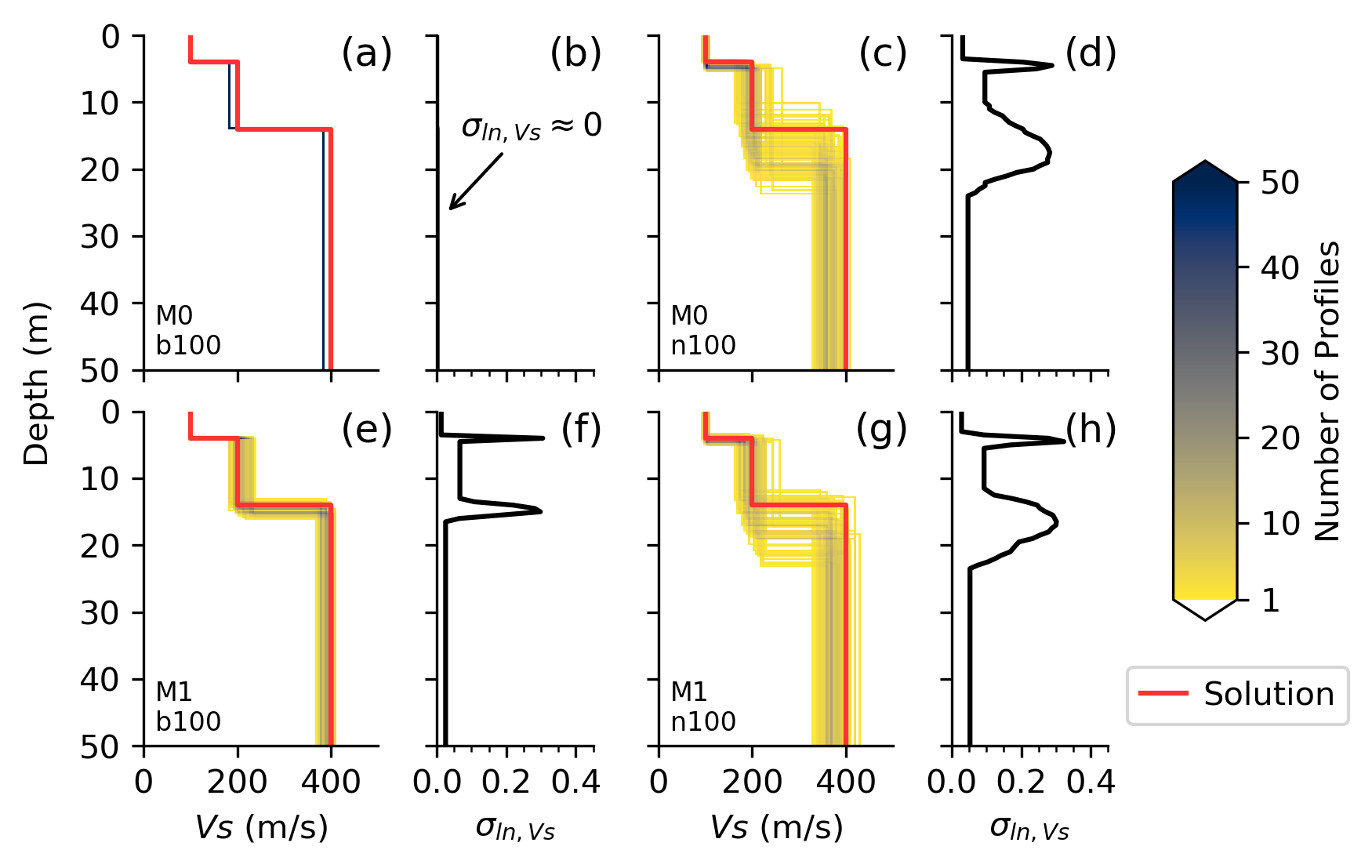}
	\caption{Shear wave velocity (Vs) profiles obtained from a synthetic experimental dispersion dataset using \textit{\textbf{four common approaches from the literature}} for accounting for Vs uncertainty in inversion. In order to better illustrate their concentration, the profiles have been discretized in terms of depth and Vs, binned into cells, and color mapped in terms of the number of profiles in each cell.  The presented approaches include: (a) a minimum misfit of 0 (M0) inversion with the selection of the 100 lowest misfit/``best'' models (b100), (c) a M0 inversion with 100 random profiles selected from all models with a misfit less than 1.0 (n100), (e) a minimum misfit of 1.0 (M1) inversion with the selection of the 100b models, and (g) an M1 inversion with the selection of the n100 models. The Vs profiles from these approaches are shown alongside the solution profile in red. Adjacent to each suite of profiles is their corresponding lognormal standard deviation of Vs ($\sigma_{ln,Vs}$). The sharp spikes in $\sigma_{ln,Vs}$ are due to the uncertainty in the profile's layer boundaries, reflecting a shortcoming in how $\sigma_{ln,Vs}$ has been calculated historically, and are not the result of actual uncertainty in Vs directly.}
	\label{fig:2}
\end{figure}

\begin{figure}[!t]
	\centering
	\includegraphics[width=3.9in]{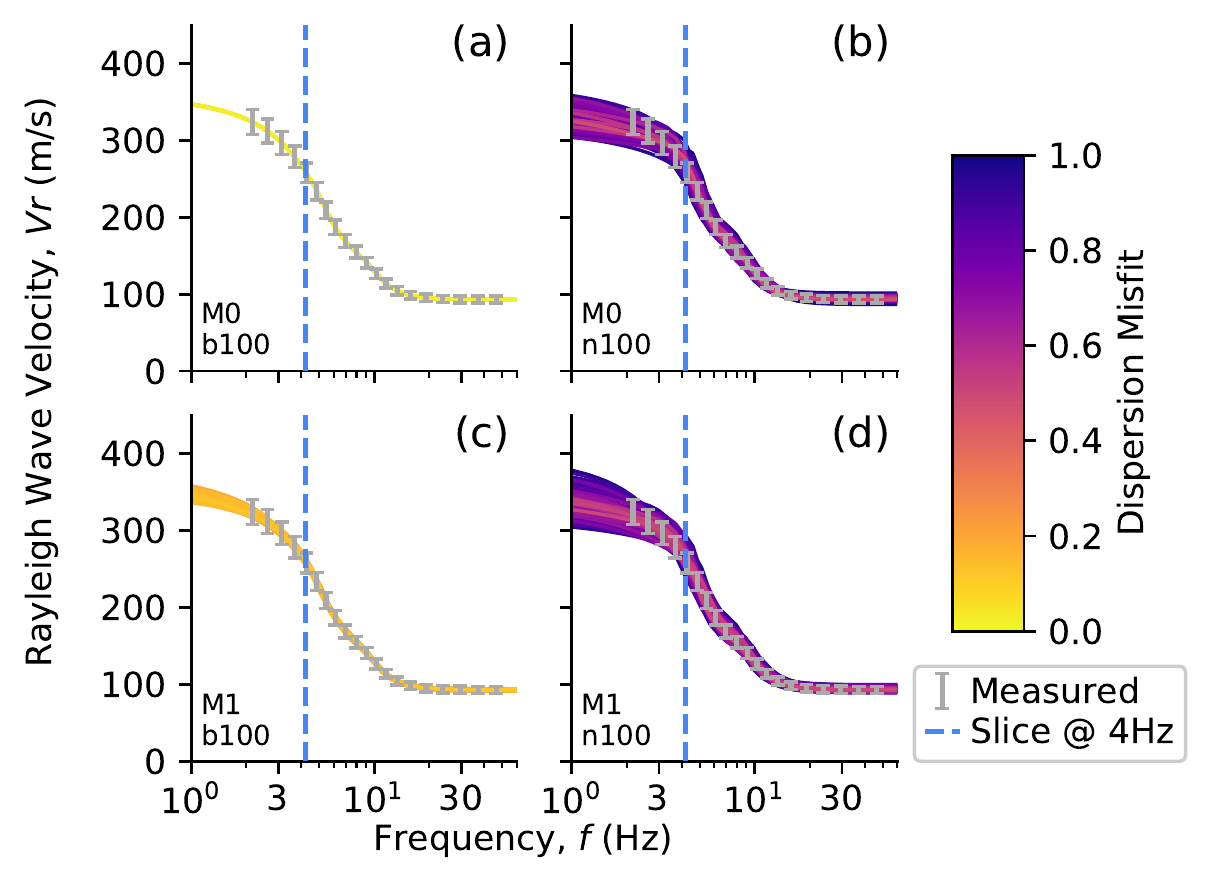}
	\caption{Comparison of the experimental dispersion data from a synthetic dataset with the individual theoretical dispersion curves obtained from inversion using \textbf{\textit{four common approaches from the literature}} for accounting for Vs uncertainty. Each of the four panels (a)-(d) illustrate one of the four common approaches from the literature. They include: (a) a minimum misfit of 0 (M0) inversion with the selection of the 100 lowest misfit/``best'' models (b100), (b) a M0 inversion with 100 random profiles selected from all models with a misfit less than 1.0 (n100), (c) a minimum misfit of 1.0 (M1) inversion with the selection of the b100 models, and (d) an M1 inversion with the selection of the n100 models. The vertical dashed blue line at approximately 4 Hz, shown on all panels, indicates the location where a ``slice'' is shown in Figure \ref{fig:4} for each of the four approaches.}
	\label{fig:3}

\end{figure}

For illustration purposes and to simplify initial discussions, the experimental dispersion data was inverted using only a single, three-layer parameterization (i.e., a Layering by Number (LN) = 3 parameterization), as this layering parameterization was shown to perform the best when inverting the example dataset in the previous study by Vantassel and Cox (\citeyear{vantassel_swinvert_2021}). The experimental dispersion data were inverted using the Neighborhood Algorithm \citep{sambridge_geophysical_1999} as implemented in the Dinver module of Geopsy \citep{wathelet_geopsy_2020}. For both the M0- and M1-style inversions, an initial 10,000 random trial models followed by 50,000 neighborhood-algorithm trial models were considered. Note that the number of neighborhood-algorithm models used here is less than those used in previous studies from the literature, which tended to use a hundred-thousand or more. We do this to mitigate bias in the resulting suite of acceptable models caused by the inclusion of many (potentially tens-of-thousands) very similar models with misfits close to zero. We believe the use of a smaller, but sufficient \citep{vantassel_swinvert_2021}, number of models provides these techniques the best possible chance of successfully propagating the experimental dispersion data's uncertainty into the resulting models.  The selected b100 and n100 models are shown in terms of their Vs profiles and associated uncertainty in Figure \ref{fig:2}. Note, to better illustrate their concentration, the Vs profiles have been discretized in terms of depth and Vs, binned into cells, and color mapped in terms of the number of profiles in each cell. Vs uncertainty is expressed in the form of the lognormal standard deviation of Vs ($\sigma_{ln,Vs}$), which is commonly used in seismic hazard studies, and is similar to the COV for values less than approximately 0.3.  The theoretical dispersion curves for the b100 and n100 models are presented with the experimental dispersion data in Figure \ref{fig:3} so the reader can view how the apparent uncertainty in the inversion-derived Vs profiles relates to the apparent uncertainty in the theoretical dispersion curves.

\begin{figure}[!t]
	\centering
	\includegraphics[width=2.9in]{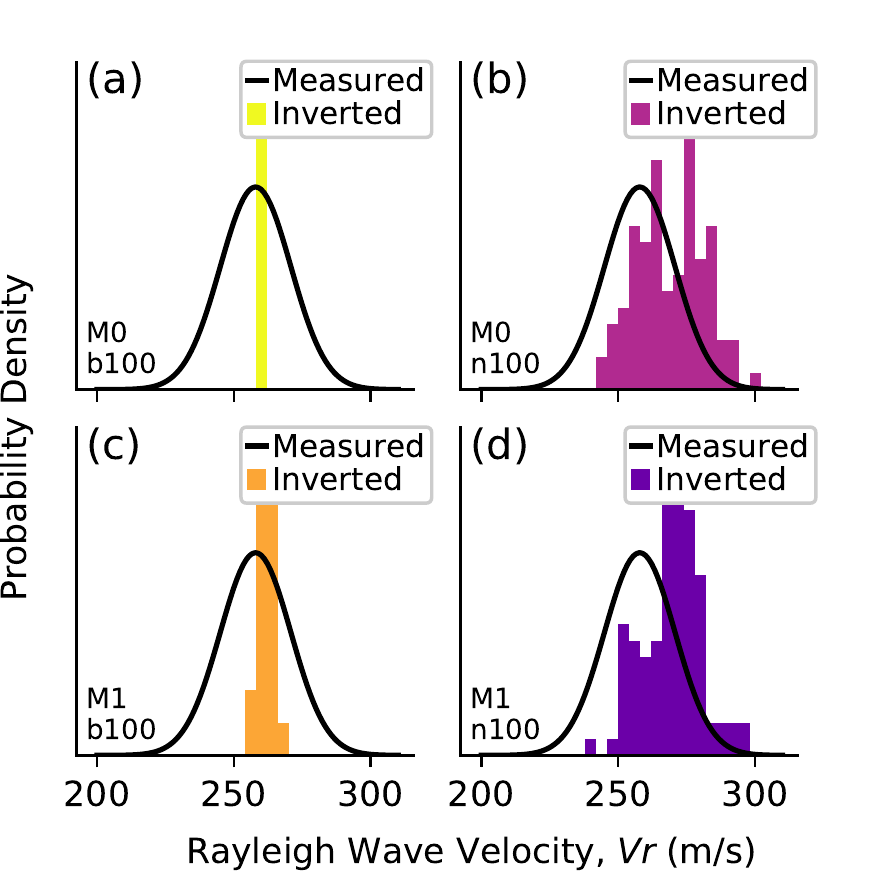}
	\caption{Distributions of the measured (i.e., experimental) and inverted (i.e., theoretical) dispersion data at a frequency of approximately 4 Hz using \textbf{\textit{four common approaches from the literature}} to account for Vs uncertainty in surface wave inversion. The four panels have the same caption as those shown in Figure \ref{fig:3}.}
	\label{fig:4}
\end{figure}

\begin{figure}[!t]
	\centering
	\includegraphics[width=4in]{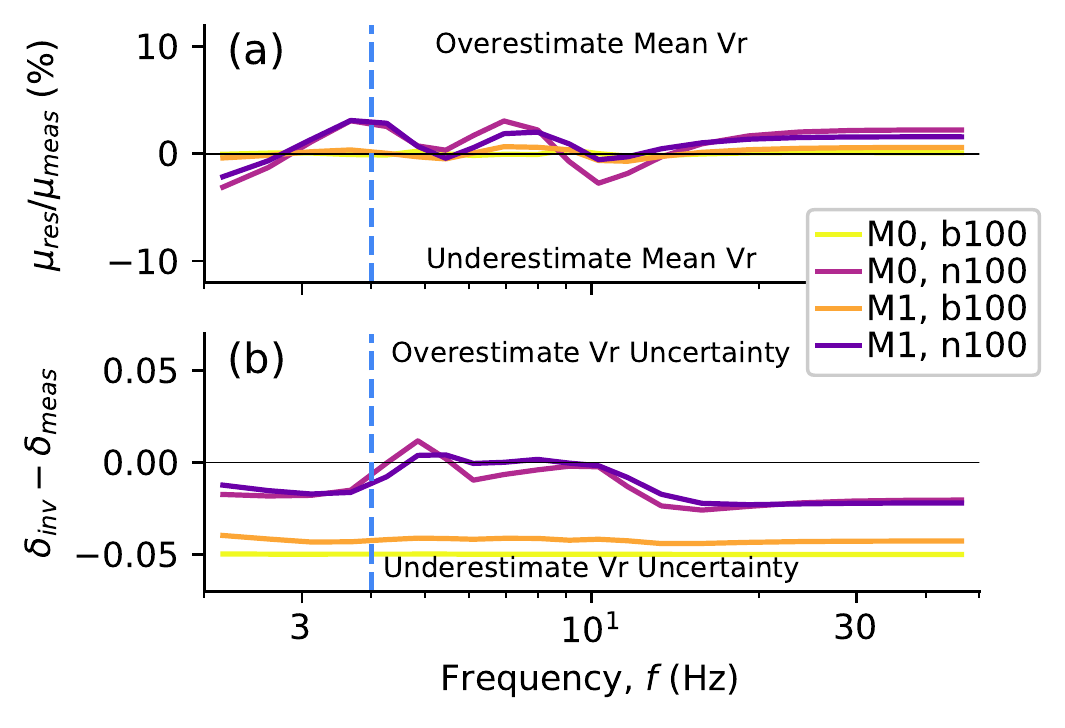}
	\caption{Quantitative assessment of \textbf{\textit{four common approaches from the literature}} for propagating uncertainty in experimental dispersion data to the inverted Vs profiles. The comparison is made on the basis of: (a) the dispersion residual mean ($\mu_{res}$) [i.e., the difference between the mean of the inverted theoretical dispersion curves ($\mu_{inv}$) and the mean of the measured experimental dispersion data ($\mu_{meas}$)] normalized by $\mu_{meas}$, expressed in percent, and (b) the dispersion residual coefficient of variation [i.e., the difference between the coefficient of variation of the inverted theoretical dispersion curves ($\delta_{inv}$) and the coefficient of variation of the measured experimental dispersion data ($\delta_{meas}$)]. The vertical dashed blue line at approximately 4 Hz in panels (a) and (b) indicates the location where a “slice” was shown in Figure \ref{fig:4} for each of the four approaches.}
	\label{fig:5}
\end{figure}

Figure \ref{fig:2} illustrates that the resulting Vs profiles are most sensitive to how they are sampled (b100 vs n100), and significantly less dependent on how the inversion is performed (M0 vs M1). This is shown both qualitatively when viewing the range of Vs profiles [i.e., panels (a), (c), (e), and (g)] and quantitatively when viewing $\sigma_{ln,Vs}$ [i.e., panels (b), (d), (f), and (h)]. For example, the Vs profiles obtained from the b100 approach show minimal scatter/uncertainty, while those obtained from the n100 approach show significantly more scatter/uncertainty.  Note that the spikes in $\sigma_{ln,Vs}$ are due to the uncertainty in the layer boundaries, reflecting a shortcoming in how $\sigma_{ln,Vs}$ has been traditionally calculated, and are not the result of uncertainty in Vs directly. The trends in Vs scatter/uncertainty shown in Figure \ref{fig:2} are tied directly to visible trends in the scatter/uncertainty of the theoretical dispersion curves shown in Figure \ref{fig:3}. The color scale in Figure \ref{fig:3} indicates the range of dispersion misfit values, which varies widely between approaches. The b100 models are primarily composed of models with low misfit dispersion misfit models (\textless 0.2) which fit the mean trend of the experimental data very well, but severely underestimate the experimental uncertainty. In contrast, the n100 models contains essentially all higher misfit models (\textgreater 0.6) that qualitatively appear to fit the experimental dispersion data's uncertainty, although some biases can be observed, particularly towards lower Vr values at low frequencies (e.g., in Figure \ref{fig:3}b at frequencies below 4 Hz). To better illustrate these problems, the vertical blue dashed line at 4 Hz in each panel of Figure \ref{fig:3} corresponds to a ``slice'' shown in each of the corresponding panels of Figure \ref{fig:4}. The mismatch between the measured and inverted distributions of Vr in Figure \ref{fig:4} clearly show that none of the methods from the literature are able to capture the mean, standard deviation, and distribution of the experimental dispersion data. To illustrate this quantitatively across the entire frequency range, Figure \ref{fig:5} shows the normalized residual mean and residual COV for the dispersion data for the four approaches considered. The vertical dashed blue line again indicates the location of the ``slices'' shown in Figure \ref{fig:4}. Figure \ref{fig:5}a confirms that using b100 sampling will tend to result in a suite of models that well-fit the mean trend of the experimental dispersion data (i.e., normalized residual mean near zero). Whereas, using n100 sampling may result in significant errors in the mean (e.g., overestimation of Vr at 4 Hz). Figure \ref{fig:5}b illustrates that using b100 sampling, which was shown to fit the mean Vr so precisely in Figure \ref{fig:5}a, severely underestimates the true dispersion uncertainty at all frequencies. Recall that the dispersion uncertainty for this synthetic example was set using a COV of 0.05 at all frequencies; meaning that the b100 sampling with a residual COV of nearly -0.05 underestimates the true dispersion uncertainty by nearly 100\%. In contrast, the n100 models that did not well-fit the mean trend tend to provide a better representation of the experimental dispersion uncertainty, although in this case they still underestimate the measured uncertainty at both high and low frequencies by approximately 0.02, or 40\%. These results clearly indicate that the currently available methods of accounting for Vs uncertainty in surface wave inversion are unable to propagate the uncertainty in the experimental dispersion data into the resulting suites of Vs profiles.

\section{A New Procedure to Account for Vs Uncertainty in Surface Wave Inversion}

\begin{figure}[!t]
	\centering
	\includegraphics[width=3.75in]{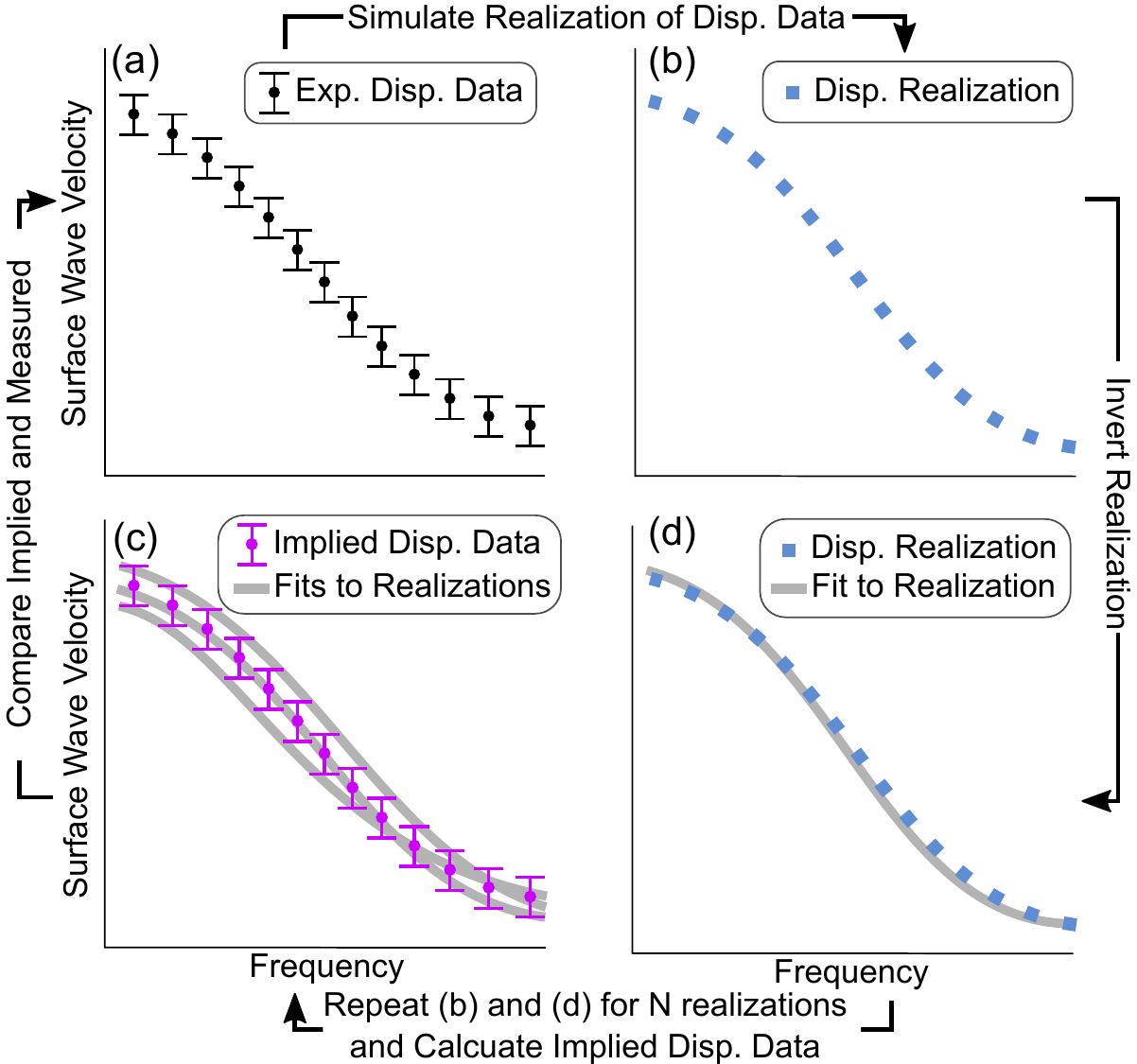}
	\caption{A new procedure for developing uncertainty-consistent shear wave velocity (Vs) profiles from surface wave dispersion data. The procedure involves: (a) describing the site's experimental dispersion data in terms of its statistics, (b) generating a realization of the experimental dispersion data, (d) inverting the realization to obtain a single, best theoretical fit, repeating (b) and (d) for N realizations, and (c) comparing the statistics of the resulting theoretical dispersion curves (i.e., the implied dispersion data) to that of the measured experimental dispersion data.}
	\label{fig:6}
\end{figure}

The four example approaches from the literature considered in the previous section had three specific shortcomings: (1) the profiles tended to capture either the mean trend or the variance of the dispersion data, but not both (refer to Figure \ref{fig:5}), (2) theoretical dispersion curves from the suites of inverted models did not follow the distribution of the synthetic experimental dispersion data (refer to Figure \ref{fig:4}), and (3) when the results were found to be unacceptable, as was the case for the synthetic example discussed above, there was no clear procedure to remedy the inconsistency. This section presents an alternative procedure to account for Vs uncertainty in surface wave inversion. The new procedure is presented schematically in Figure \ref{fig:6}. Figure \ref{fig:6}a shows experimental dispersion data with site-specific measurements of uncertainty. This uncertainty is quantified in terms of surface wave phase velocity at each frequency using a mean, standard deviation, and, very importantly, correlation coefficients between all frequency pairs. All of these statistics can be easily quantified by the analyst using the approach to developing dispersion data with site-specific uncertainty discussed previously (recall Figure \ref{fig:1}). Consider the experimental dispersion data in Figure \ref{fig:6}a, which contains 13 frequency points. Its statistics are completely described using 13 mean values (shown with a circle), 13 standard deviations (shown with error bars), and a matrix of 13x13 correlation coefficients (not shown) relating the phase velocity at each frequency to the other 12 values. With the experimental dispersion data described in terms of its statistics, it's now possible to simulate a realization from that experimental dispersion data, shown schematically in Figure \ref{fig:6}b. It is during this simulation stage that the inclusion of the correlation information is so critical. Without such information one is left to assume the correlations, or worse, independence.
However, doing so would likely result in simulated dispersion curves that were erratic from frequency-to-frequency and inconsistent with the dispersion data that was used to estimate the uncertainty. In essence, the inclusion of the correlations between dispersion points imposes a relational constraint that helps to encourage simulated dispersion curves that are consistent with the measured data (i.e., in this case, smooth and continuous). Once a curve has been simulated, it is inverted to obtain a single, ``best'' fit model, as shown in Figure \ref{fig:6}d. Note that while all previous procedures to account for Vs uncertainty required the use of a global-search algorithm to ultimately develop suites of Vs profiles, in contrast, this new procedure is inversion-algorithm agnostic, thereby permitting the analyst to use any algorithm of their choice (global or local). Furthermore, by selecting only the single, ``best'' model the issue of dependence on the number of trial models is avoided altogether, provided of course that a reasonable/sufficient number of models has been attempted to produce a good fit to the simulated dispersion curve, which is easily verifiable. Returning to Figure \ref{fig:6}d, it is important to note that the fit may not be perfect, however, the proposed procedure does not require it. The issue of primary importance is that the simulated experimental dispersion data and the theoretical fit are in good agreement over the experimental frequency range, though not necessarily strictly identical. The procedure of simulating a realization of the dispersion data and fitting it through inversion is repeated for some number of trials (N), which is the only input parameter to be defined by the analyst. The best inversion-derived fit to each realization and the corresponding ground model is saved as a potential solution to the experimental data considering its uncertainties, refer to Figure \ref{fig:6}c. The N fits (i.e., N theoretical dispersion curves) can then be used to calculate implied dispersion data statistics which can be compared directly against the experimental dispersion data statistics to assess the successfulness of the procedure. Note that the value of N is expected to be problem dependent, however, a minimum acceptable value for N can be checked prior to performing any inversions by comparing the statistics of the simulated dispersion realizations and the experimental dispersion data. If the statistics of the simulated dispersion data (i.e., the targets used for inversion) are unable to reproduce the experimental statistics, it can then be inferred that the inversion-derived fits to those simulated curves will also not reproduce the experimental statistics. Thus, care must be taken to select N to be sufficiently large to ensure agreement between the simulated and measured experimental data, although not too large to avoid excessive computational cost when performing N different inversions. An N=250 was shown to perform well over the course of this study, however the choice of N is left to the discretion of the analyst. Importantly, following the inversion of the N simulated dispersion curves to obtain N theoretical dispersion curves, the statistics implied by the N theoretical dispersion curves (i.e., the implied dispersion data, refer to Figure \ref{fig:6}c), must be compared quantitatively with the measured experimental dispersion data in a manner similar to that shown in Figure \ref{fig:5}. If the agreement between the implied and measured experimental dispersion data is found to be unsatisfactory, additional simulations and/or increasing the value of N can be used to improve the results. The remainder of this paper will address the application of this new approach to two synthetic tests and a real-world example.

\section{Synthetic Tests}

\subsection{For a Single Layering Parameterization}

To demonstrate the effectiveness of the newly proposed procedure, we will test it using the same synthetic dataset as that was used previously to evaluate the literature-based approaches for accounting for Vs uncertainty. Recall, the experimental dispersion data (i.e., mean and standard deviation as a function of frequency) for this dataset is presented in Figure \ref{fig:1}c. As the new approach requires the correlations between frequencies, and the synthetic data, as developed, does not include such information, the correlations had to be synthesized. The procedure for synthesizing the correlations involved first simulating ground models whose theoretical dispersion curves were consistent with the experimental dispersion's uncertainty. The statistics used to inform the simulation of these ground models were developed based on the mean values from an M0-type inversion and the uncertainty from an M1-type inversion, and while these types of inversions have been shown to be lacking in their ability to propagate dispersion uncertainty (refer to Figures \ref{fig:4} and \ref{fig:5}), they provided a reasonable correlation structure. The synthesized correlations were combined with the known/assumed statistics (refer to Figure \ref{fig:1}c) to define the experimental dispersion data. With this information, the procedure outlined in Figure \ref{fig:6} was performed using N=250 realizations. The choice of N=250 was made prior to performing the inversions by checking that the statistics of the realizations consistently reproduced the statistics of the experimental dispersion data. Each realization was inverted using the same 3-layer parameterization and 10,000 random plus 50,000 neighborhood algorithm trial models, as was used in the previous section. Inversions were performed on the Texas Advanced Computing Center's (TACCs) cluster Stampede2 using a single Skylake (SKX) node. The entire analysis (i.e., the inversion of all N=250 simulated dispersion curves) took less than 2 hours to complete.

\begin{figure}[!t]
	\centering
	\includegraphics[width=5in]{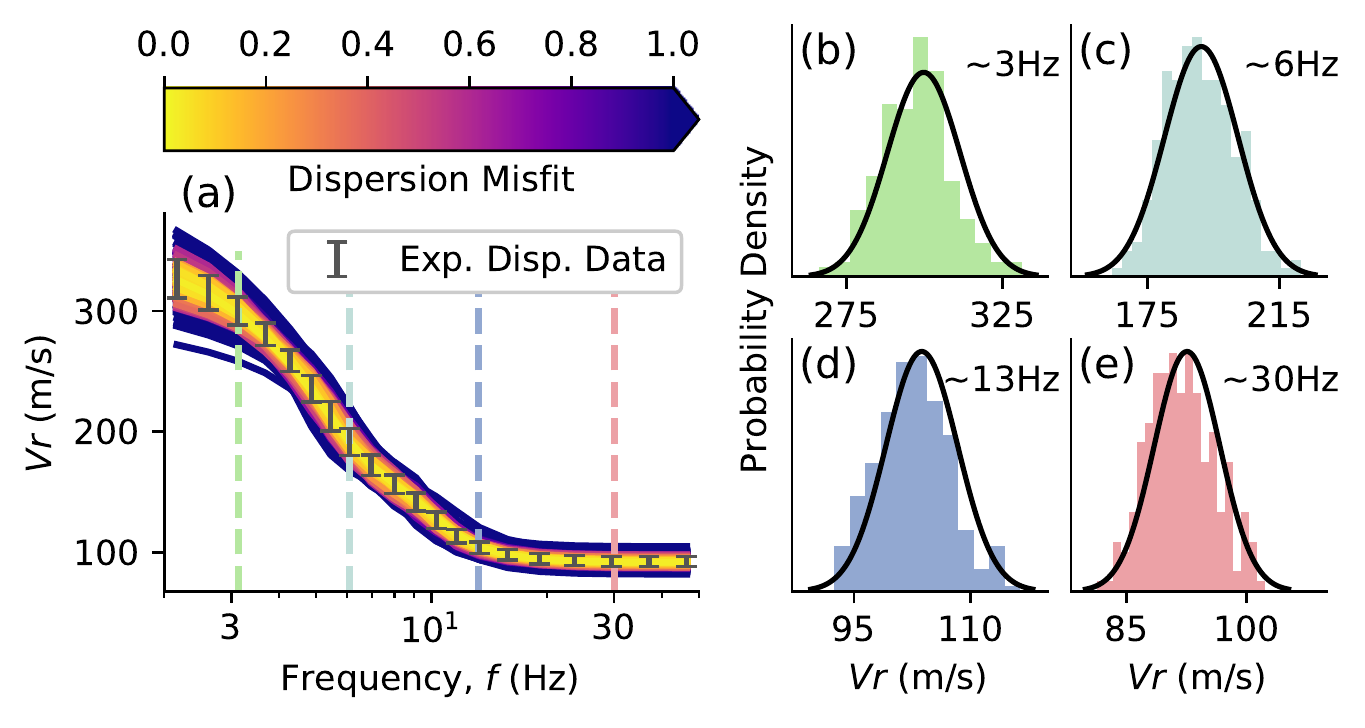}
	\caption{Qualitative assessment of the \textbf{\textit{newly proposed procedure}} for propagating experimental dispersion uncertainty into the inverted Vs profiles \textbf{\textit{considering a single inversion parameterization for a synthetic example}}. Panel (a) shows the experimental dispersion data and 250 inverted theoretical dispersion curves fit to the 250 realizations of the experimental dispersion data. The theoretical dispersion curves have been colored according to their dispersion misfit values relative to the original experimental dispersion data, not their respective realizations. The vertical dashed lines in panel (a) at approximately 3, 6, 13, and 30 Hz denote the location of the ``slices'' shown in panels (b), (c), (d), and (e), respectively. These ``slices'' compare the measured experimental dispersion data's distribution (solid black line) with the distribution of Rayleigh wave velocity (Vr) derived from inversion (histogram).}
	\label{fig:7}
\end{figure}

\begin{figure}[!t]
	\centering
	\includegraphics[width=4in]{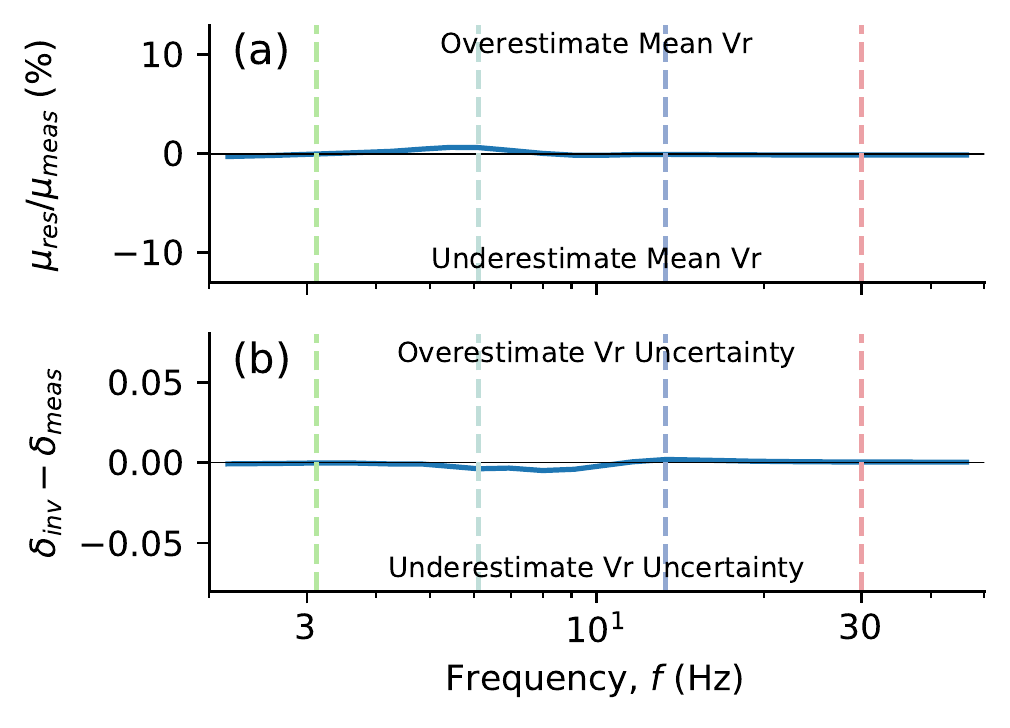}
	\caption{Quantitative assessment of the \textbf{\textit{newly proposed procedure}} for propagating experimental dispersion uncertainty into the inverted Vs profiles \textbf{\textit{considering a single inversion parameterization for a synthetic example}}. The comparison is made on the basis of: (a) the dispersion residual mean ($\mu_{res}$) [i.e., the difference between the mean of the inverted theoretical dispersion curves ($\mu_{inv}$) and the mean of the measured experimental dispersion data ($\mu_{meas}$)] normalized by $\mu_{meas}$ and expressed in percent, (b) the dispersion residual coefficient of variation [i.e., the difference between the coefficient of variation of the inverted theoretical dispersion curves ($\delta_{inv}$) and the coefficient of variation of the measured experimental dispersion data ($\delta_{meas}$)]. The vertical dashed lines in panels (a) and (b) denote the location of the ``slices'' shown in Figure \ref{fig:7}.}
	\label{fig:8}
\end{figure}

\begin{figure}[!t]
	\centering
	\includegraphics{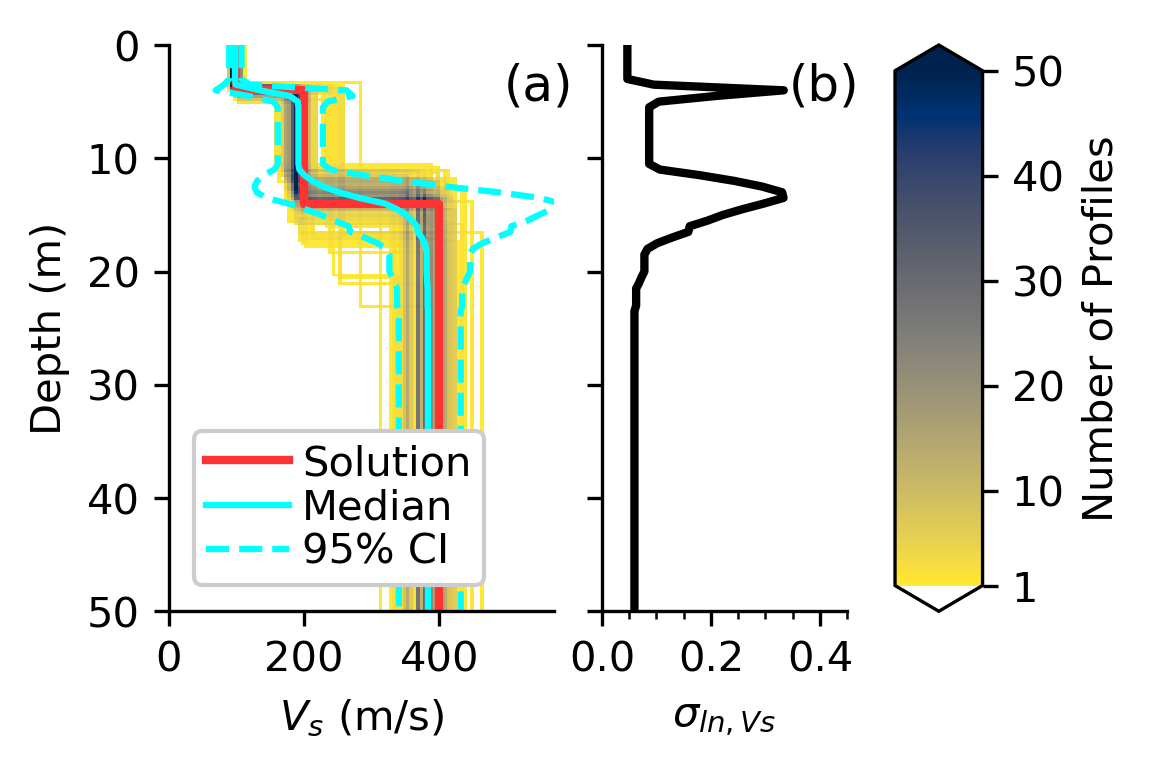}
	\caption{Uncertainty-consistent Vs profiles \textbf{\textit{considering a single inversion parameterization for a synthetic example}} using the \textbf{\textit{newly proposed procedure}}. In order to better illustrate their concentration, the profiles have been discretized in terms of depth and Vs, binned into cells, and color mapped in terms of the number of profiles in each cell. Panel (a) summarizes the Vs profiles resulting from the inversion of the N=250 realizations of the experimental dispersion data alongside the true solution, the discretized log-normal median, and the 95\% confidence interval (CI). Panel (b) illustrates the lognormal standard deviation of Vs ($\sigma_{ln,Vs}$) of the 250 profiles shown in panel (a). The sharp spikes in $\sigma_{ln,Vs}$ are due to the uncertainty in the profile's layer boundaries, reflecting a shortcoming in how $\sigma_{ln,Vs}$ has been calculated historically, and are not the result of actual uncertainty in Vs directly.}
	\label{fig:9}
\end{figure}

The results from the inversion analyses are provided in Figures \ref{fig:7}, \ref{fig:8}, and \ref{fig:9}. Figure \ref{fig:7}a compares the theoretical dispersion curves fit to the N=250 realizations of the experimental dispersion data with the original experimental dispersion data. Note the theoretical dispersion curves have been colored in terms of their dispersion misfit relative to the original experimental data and not their respective realizations. In contrast to the previous methods (refer to Figure \ref{fig:3}) we observed a reasonable mixture of theoretical dispersion curves with low misfits (i.e., \textless 0.2) near the mean trend of the experimental data and higher misfits extending outward to capture the dispersion data's uncertainty. This indicates qualitatively that the theoretical curves from the new procedure are being distributed according to the uncertainty in the experimental dispersion data rather than directly as a consequence of the inversion process. To better illustrate the agreement between the measured and inverted dispersion uncertainty, the vertical dashed lines at approximately 3, 6, 13, and 30 Hz in Figure \ref{fig:7}a indicate the locations of ``slices'' shown in Figures \ref{fig:7}b, c, d, and e, respectively. These ``slices'' indicate excellent agreement between the measured dispersion data's probability density functions (black lines) and the inverted distributions of Vr (histograms), especially in comparison to the previous methods considered (refer to Figure \ref{fig:4}). Figure \ref{fig:8} presents the results in a quantitative manner across the entire frequency range. Figures \ref{fig:8}a and b, present the normalized dispersion residual mean and dispersion residual COV, respectively. While minor differences are observed, both metrics illustrate excellent agreement between the experimental and inverted dispersion data's mean and uncertainty (i.e., residuals approximately equal to zero at all frequencies), especially when compared to previous methods (refer to Figure \ref{fig:5}).

With the theoretical dispersion curves from the inverted ground models having been shown to be consistent with the uncertainty of the experimental dispersion data, we can now examine the effects on the Vs profiles. Figure \ref{fig:9}a presents the single-lowest misfit Vs profile from the inversion of each of the N=250 dispersion realizations. To better illustrate their concentration about the true solution the Vs profiles have been discretized in terms of depth and Vs, binned into cells, and color mapped in terms of the number of profiles in each cell. The discretized lognormal median Vs profile and the 95\% lognormal confidence interval (CI) profiles are also shown for reference. Figure \ref{fig:9}a shows that the discretized median profile reasonably captures the true solution and that the 95\% confidence interval qualitatively captures the variance in the Vs profiles, with the notable exception of the sharp spikes at layer boundaries. When we compare $\sigma_{ln,Vs}$ from the new approach (i.e., Figure \ref{fig:9}b) directly with those from the literature (i.e., Figure \ref{fig:2}b, d, f, and h) we observe that the new approach shows increased uncertainty from the (b100) alternatives, as should be expected, but surprisingly slightly less uncertainty than the (n100) alternatives. This observation indicates that (at least for this example) the n100 models tend to over-estimate Vs uncertainty even when under-predicting the dispersion data uncertainty (refer to Figure \ref{fig:5}). Further examination of Figure \ref{fig:9}b shows that  $\sigma_{ln,Vs}$ for this example is approximately 0.06, which is quite low compared to values commonly assumed in practice \citep{epri_seismic_2012, stewart_guidelines_2014, toro_probabilistic_1995}. This is certainly due in part to the use of only a single inversion layering parameterization, as others have shown that the variability within a parameterization is generally much less than that between parameterizations. Hence, if the true subsurface layering is unknown a priori, it is not simply enough to accurately represent the experimental dispersion data's uncertainty using a single, assumed layering parameterization. Rather, one must also incorporate the epistemic uncertainty in the layering parameterization itself \citep{cox_layering_2016, di_giulio_exploring_2012, vantassel_swinvert_2021}. The incorporation of multiple inversion parameterizations into the procedure and their effects on the Vs profiles are presented in the following section. 

\subsection{Extension to Multiple Inversion Layering Parameterizations}

Uncertainty within surface wave inversion is generally split into intra- and inter-parameterization variability, which address the uncertainty inside a single inversion parameterization (i.e., the previous example) and between various parameterizations (i.e., current example), respectively. Previous studies have shown that the inter-parameterization variability is generally more significant than the inter-parametrization variability. However, as the methods used in previous studies to account for intra-parameterization variability tended to underestimate the experimental uncertainty (refer to Figures \ref{fig:4} and \ref{fig:5}), it is of interest to reexamine this conjecture using the new procedure.

The new procedure was repeated using the same synthetic experimental dispersion data as the previous example (refer to Figure \ref{fig:1}c), except rather than using only a single layering parameterization consisting of 3 layers, five different layering parameterizations were used to account for epistemic uncertainty in the parameterization selection. The five parameterizations were based on the Layering by Number (LN) schema and included LNs of 3, 5, 7, 9, and 14. These parameterizations were selected because they were all deemed acceptable when this same set of experimental dispersion data was inverted by Vantassel and Cox (\citeyear{vantassel_swinvert_2021}). In practice, the analyst must decide carefully which parameterizations to pursue, as this can strongly impact the resulting inversion-derived Vs profiles. We recommend investigating a number of trial parameterizations, and the reader is referred to the previous study by Vantassel and Cox (\citeyear{vantassel_swinvert_2021}) for a more thorough discussion on the number and type of parameterizations to use in their inversions. The five parameterizations noted above were used to invert different sets of N=250 dispersion realizations (i.e., 1250 realizations in total). N=250 realizations per parameterization was selected to ensure that each parameterization would have a sufficient number of realizations to each individually capture the uncertainty of the experimental dispersion data, as it is important to not discount the intra-parameterization variability when investigating the inter-parameterization variability. This, of course, is not the only approach available to consider the inter-parameterization variability. An alternative method examined during the course of this study is to invert a single set of N realizations with various parameterizations. This alternate approach was found to produce similar results to those using the presented method, however, since the presented method is believed to be more robust, as it ensures each parameterization remains separate, it is the one selected here.

\begin{figure}[!t]
	\centering
	\includegraphics[width=5.5in]{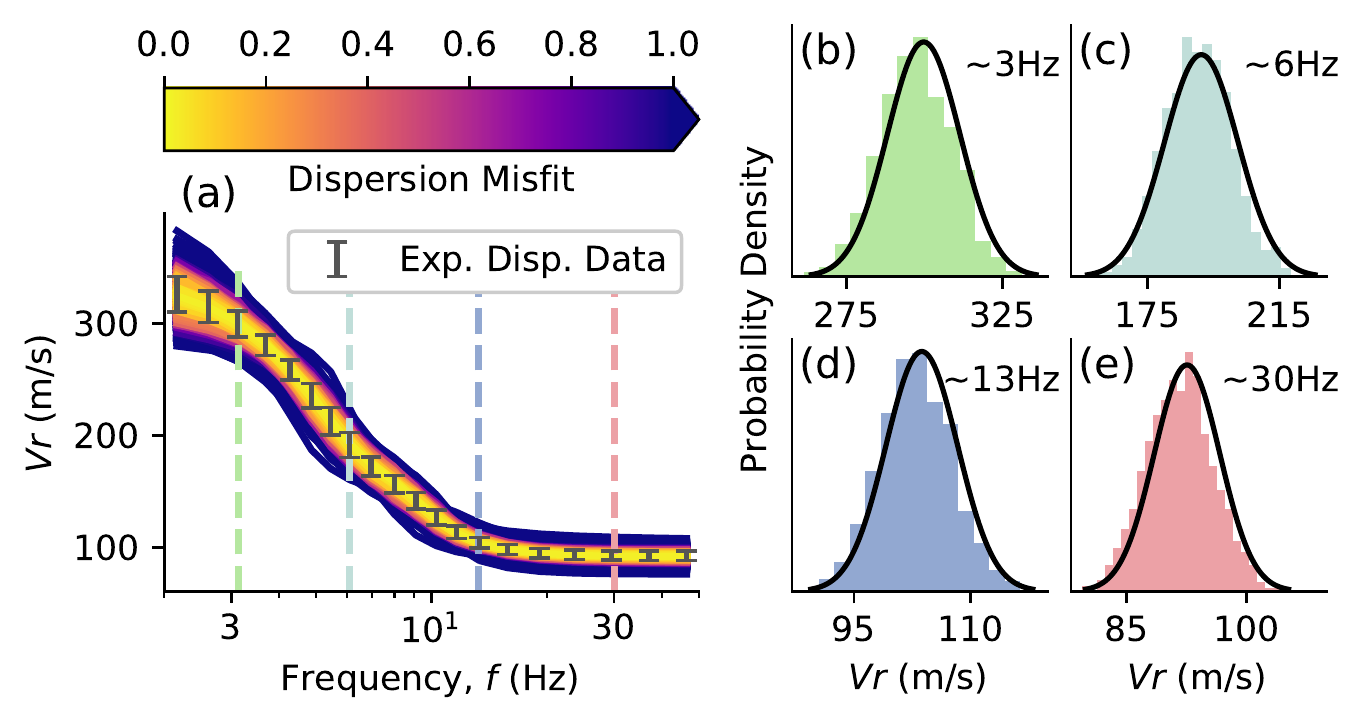}
	\caption{Qualitative assessment of the \textbf{\textit{newly proposed procedure}} for propagating experimental dispersion uncertainty into the inverted Vs profiles \textbf{\textit{considering multiple inversion parameterizations for a synthetic example}}. Panel (a) shows the experimental dispersion data and the inverted theoretical dispersion curves fit to the 1250 realizations (5 parameterizations with N=250 realizations each) of the experimental dispersion data. The theoretical dispersion curves have been colored according to their dispersion misfit values relative to the original experimental dispersion data, not their respective realizations. The vertical dashed lines in panel (a) at approximately 3, 6, 13, and 30 Hz denote the location of the “slices” shown in panels (b), (c), (d), and (e), respectively. These “slices” compare the measured experimental dispersion data's distribution (solid black line) with the distribution of Rayleigh wave velocity (Vr) derived from inversion (histogram).}
	\label{fig:10}
\end{figure}

\begin{figure}[!t]
	\centering
	\includegraphics[width=3.75in]{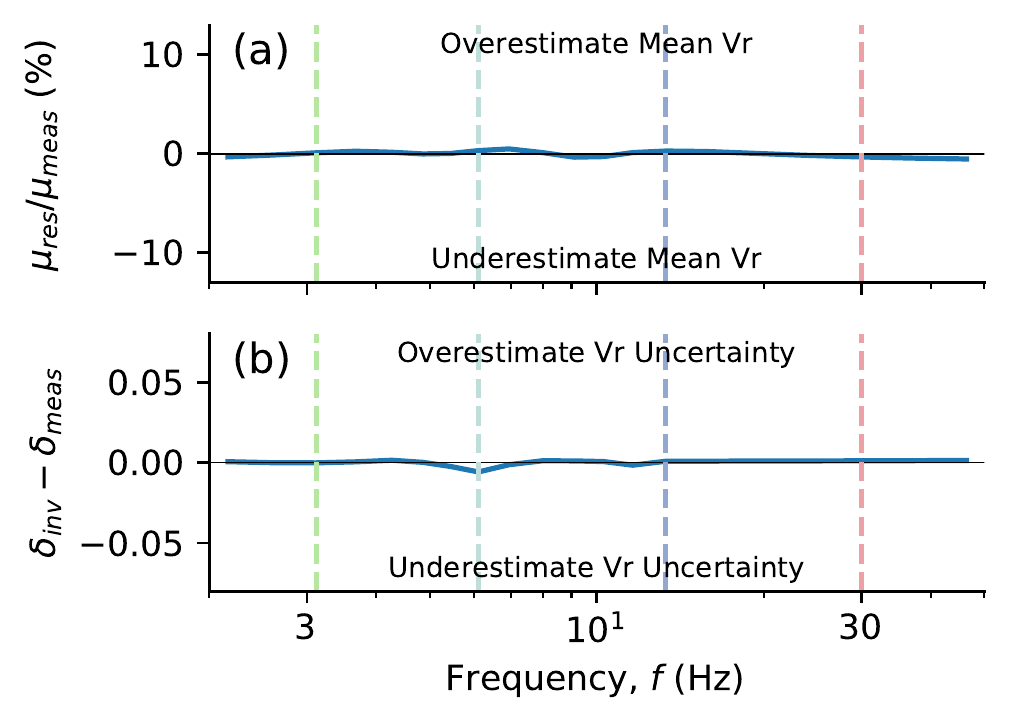}
	\caption{Quantitative assessment of the \textbf{\textit{newly proposed procedure}} for propagating experimental dispersion uncertainty into the inverted Vs profiles \textbf{\textit{considering multiple inversion parameterization for a synthetic example}}. The comparison is made on the basis of: (a) the dispersion residual mean ($\mu_{res}$) [i.e., the difference between the mean of the inverted theoretical dispersion curves ($\mu_{inv}$) and the mean of the measured experimental dispersion data ($\mu_{meas}$)] normalized by $\mu_{meas}$ and expressed in percent, (b) the dispersion residual coefficient of variation [i.e., the difference between the coefficient of variation of the inverted theoretical dispersion curves ($\delta_{inv}$) and the coefficient of variation of the measured experimental dispersion data ($\delta_{meas}$)]. The vertical dashed lines in panels (a) and (b) denote the location of the ``slices'' shown in Figure \ref{fig:10}.} 
	\label{fig:11}
\end{figure}

\begin{figure}[!t]
	\centering
	\includegraphics{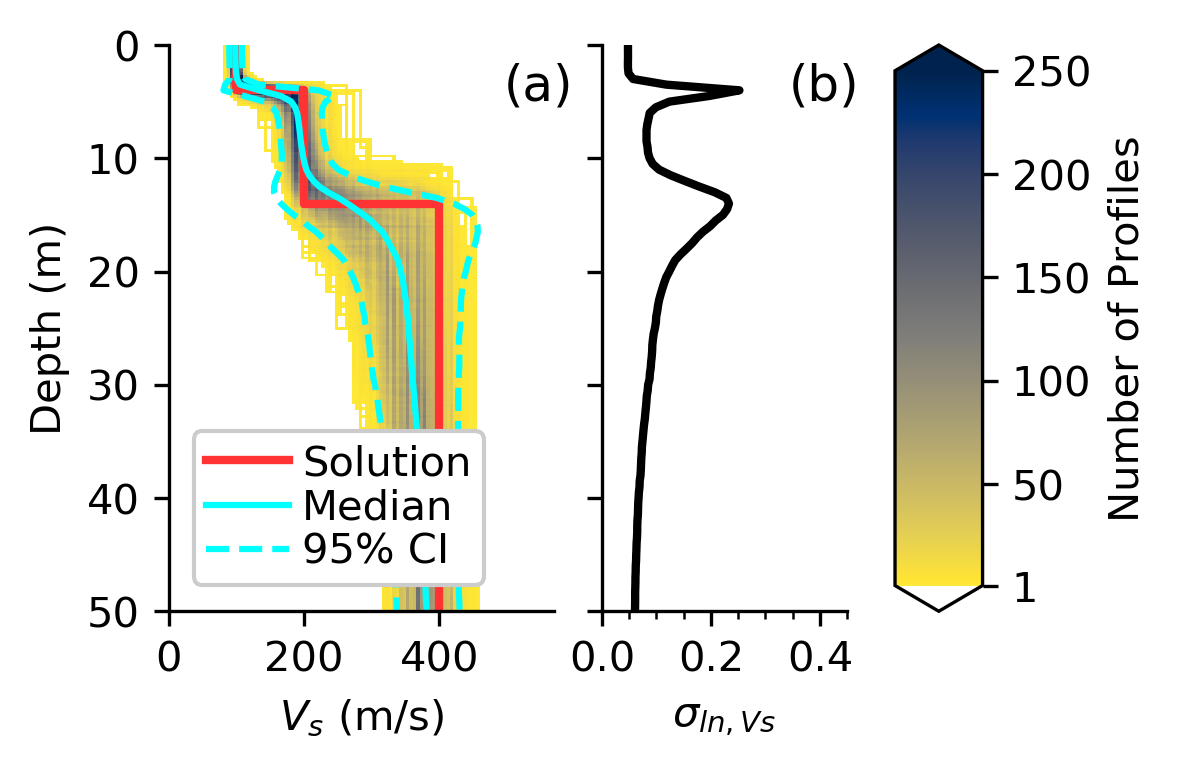}
	\caption{Uncertainty-consistent Vs profiles \textbf{\textit{considering multiple inversion parameterization for a synthetic example}} using the \textbf{\textit{newly proposed procedure}}. In order to better illustrate their concentration, the profiles have been discretized in terms of depth and Vs, binned into cells, and color mapped in terms of the number of profiles in each cell. Panel (a) summarizes the Vs profiles resulting from the inversion of 1250 realizations (5 parameterizations with N=250 realizations each) of the experimental dispersion data alongside the true solution, the discretized log-normal median, and the 95\% confidence interval (CI). Panel (b) illustrates the lognormal standard deviation of Vs ($\sigma_{ln,Vs}$) of the 250 profiles shown in panel (a). The sharp spikes in $\sigma_{ln,Vs}$ are due to the uncertainty in the profile's layer boundaries, reflecting a shortcoming in how $\sigma_{ln,Vs}$ has been calculated historically, and are not the result of actual uncertainty in Vs directly.}
	\label{fig:12}
\end{figure}

The results from the newly proposed procedure when accounting for multiple layering parameterizations are shown in Figures \ref{fig:10}, \ref{fig:11}, and \ref{fig:12}. Figure \ref{fig:10}a compares the theoretical dispersion curves fit to the 1250 realizations of the experimental dispersion data with the original experimental dispersion data. The theoretical dispersion curves are colored in terms of their dispersion misfit relative to the original experimental dispersion data (not their respective realizations) and are again shown to follow a reasonable distribution of low misfit (\textless 0.2) curves near the mean trend of the experimental data and higher misfit curves capturing its uncertainty. The vertical colored dashed lines at approximately 3, 6, 13, and 30 Hz in Figure \ref{fig:10}a indicate the locations of ``slices'' presented in Figures \ref{fig:10}b, c, d, and e, respectively. These slices reveal excellent agreement between the measured and inverted distributions of Vr. To quantitatively assess performance over the entire frequency range, Figures \ref{fig:11}a and b present the normalized dispersion residual mean and dispersion residual COV, respectively. While again some minor fluctuations are observed (e.g., at 6 Hz), both metrics illustrate excellent agreement between the experimental dispersion data's mean and uncertainty (i.e., residuals approximately equal to zero at all frequencies).

With the theoretical dispersion curves from the inverted ground models having been shown to be consistent with the uncertainty of the experimental dispersion data, we can now examine the effects on the Vs profiles. Figure \ref{fig:12}a presents the single-lowest misfit Vs profile from the inversion of each of the 1250 dispersion realizations (i.e., N=250 profiles per parameterization * 5 parameterizations = 1250 profiles). To better illustrate their concentration, the Vs profiles have been discretized in terms of depth and Vs, binned into cells, and color mapped in terms of the number of profiles in each cell. The inverted Vs profiles are shown alongside the true solution, discretized lognormal median Vs profile, and lognormal 95\% CI profiles. As expected, the inverted profiles qualitatively show more uncertainty than when a single parameterization is considered (refer to Figure \ref{fig:9}a for comparison). This distinction is most apparent at the layer boundaries, as the locations of these boundaries are controlled predominantly by the assumed layering parameterization. Parameterizations with many layers will tend to result in smoother profiles with gradual changes in Vs, while parameterizations with only a few layers will tend to result in profiles with sharper contrasts \citep{cox_layering_2016}. As the true site layering may not be known prior to surface wave inversion, there is often a need to consider multiple parameterizations with different numbers of layers to properly address one of surface wave inversion's main sources of epistemic uncertainty. Figure \ref{fig:12}b presents $\sigma_{ln,Vs}$ of the 1250 profiles. The uncertainty across multiple parameterizations is shown to increase in regards to what was observed previously for a single parameterization (refer to Figure \ref{fig:9}b for comparison). Specifically, $\sigma_{ln,Vs}$ is now closer to 0.1 within any given layer, thereby confirming quantitatively what was already observed qualitatively in Figure \ref{fig:12}a. The increase in $\sigma_{ln,Vs}$ lends confidence to the conjecture that the uncertainty resulting from multiple parameterizations will tend to exceed that for a single parameterization, although perhaps not to the same extent as may have been inferred previously due the tendencies to underestimate intra-parameterization variability in previous studies. Regardless, at least for this synthetic example, $\sigma_{ln,Vs}$ remains quite low compared to values commonly assumed in practice \citep{epri_seismic_2012, stewart_guidelines_2014, toro_probabilistic_1995}.

\section{Application at a Real Site}

\begin{figure}[!t]
	\centering
	\includegraphics[width=\textwidth]{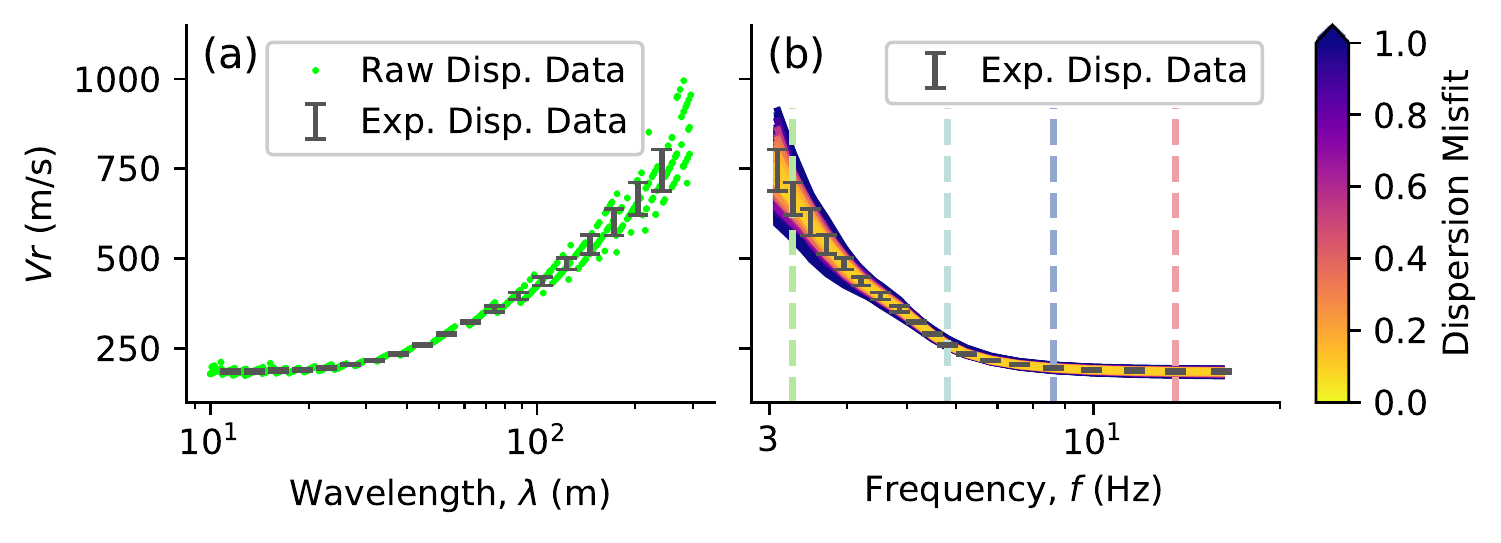}
	\caption{Dispersion data used to study the application of the \textbf{\textit{newly proposed procedure}} for propagating experimental dispersion uncertainty into the inverted Vs profiles \textbf{\textit{considering multiple parameterizations at a real site}}: (a) raw experimental dispersion data with its binned and resampled statistical representation, and (b) agreement between the experimental dispersion data and  1000 inverted theoretical dispersion curves fit to the 1000 realizations (4 acceptable parameterizations with N=250 realizations each) of the experimental dispersion data. The theoretical dispersion curves have been colored according to their dispersion misfit values relative to the original experimental dispersion data, not their respective realizations. The vertical dashed lines in panel (b) indicate the location of the ``slices'' shown in Figure \ref{fig:14}.}
	\label{fig:13}
\end{figure}

To illustrate the application of the newly proposed method at a real site, we present the following case study. For reference, the site is located in the southern California and has a known geology consisting of alluvial soils (sands and clays) approximately 15 -- 25 m thick, overlaying highly variable weathered bedrock, over competent bedrock. While an extensive set of active-source and passive-wavefield surface wave measurements are available at this site, to keep the example simple, we will only describe how dispersion data was extracted from some of the passive-wavefield microtremor array measurements (MAM). To incorporate aleatory variability across the 70 m by 70 m site into the estimates of dispersion uncertainty, four 35 m diameter circular arrays each composed of nine sensors will be considered. In essence, each of the four MAM arrays were responsible for characterizing one quarter of the overall site. While using only one size of MAM array will result in somewhat bandlimited dispersion data, we do so herein to maintain clarity in how the combined aleatory and epistemic uncertainties are estimated. MAM array sensors were buried and left undisturbed to record ambient noise, seven hours of which were used to extract dispersion data from each array. Each of the four MAM arrays were processed separately to obtain experimental dispersion data using the Rayleigh three-component beamforming method developed by Wathelet et al. (\citeyear{wathelet_rayleigh_2018}) and implemented in the open-source software Geopsy \citep{wathelet_geopsy_2020}. Constant length time blocks were preferred over frequency-dependent time blocks to facilitate statistical calculations on the experimental dispersion data. Time blocks were selected to be 60 seconds long to ensure at least 100 cycles at the lowest processing frequency. The recommended 4 time blocks per sensor per block set were used to compute the average cross-correlation matrices. As sufficiently long noise records were available, no overlap was permitted between blocks or block sets. No higher modes were apparent in preliminary dispersion processing, so only the spectral peaks with the single highest power at each frequency were selected from each block set to represent the experimental dispersion data. This resulted in 11 estimates of Vr at each frequency for each array. Or, in other words, 44 estimates of Vr per frequency across the entire site. The experimental data was binned and resampled in terms of log-wavelength following the recommendations of Vantassel and Cox (\citeyear{vantassel_swinvert_2021}). The raw experimental dispersion data and the resulting statistical representation is shown in Figure \ref{fig:13}a. The frequency-dependent COVs tend to increase from approximately 0.02 at short wavelengths to approximately 0.08 at long wavelengths. The minimum and maximum wavelength of the experimental dispersion data are approximately 10 m and 250 m, respectively. The reader will note that the maximum wavelength (250 m) is substantially longer than what one would generally expect to be able to resolve based on the array resolution limits (typically 2 or 3 times the maximum array aperture). However, this was done only after confirming that the dispersion data from these arrays were consistent with data from larger aperture arrays also performed at the site (not shown). The minimum depth of profiling can be approximated by dividing the minimum wavelength by a depth factor (df) of 3 or 2 \citep{foti_guidelines_2018} to obtain an estimate of the minimum resolvable thickness of the near-surface layer between 3 m and 5 m. These relatively large near-surface layers are the direct result of using only passive arrays of medium size for this example. The maximum depth of profiling can be approximated by dividing the maximum wavelength by a df of 3 or 2. However, the use of even a df of 3 may be optimistic when the dispersion curve has an ``L''-shape which does not flatten at long wavelengths \citep{vantassel_swinvert_2021}.  Nonetheless, we adopt a df of 3 for parameterizing the inversion's maximum depth based on the experimental dispersion data, and then limit the depth to which the inverted Vs profiles are presented based on the quality of the resulting Vs profiles (i.e., depth limited based on high $\sigma_{ln,Vs}$ values when resolution is poor).

\begin{figure}[!t]
	\centering
	\includegraphics[width=6in]{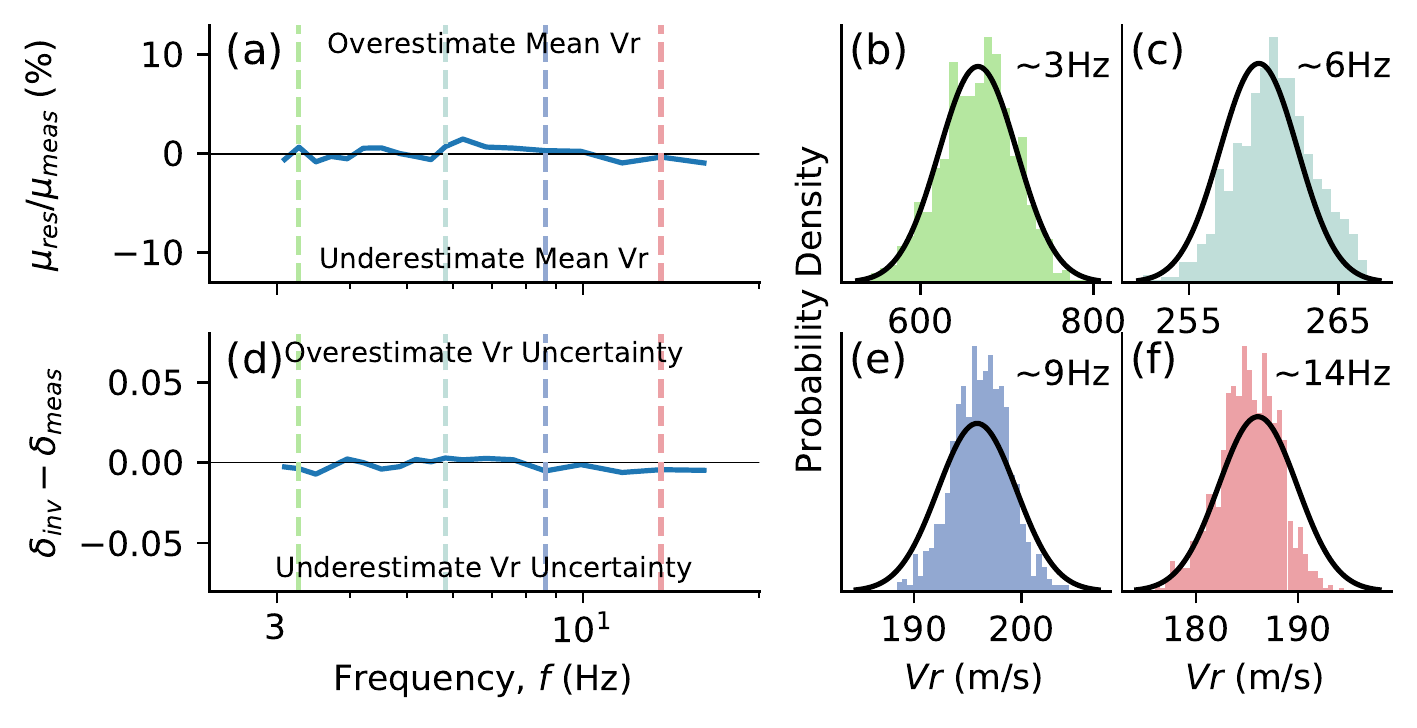}
	\caption{Quantitative assessment of the \textbf{\textit{newly proposed procedure}} for propagating experimental dispersion uncertainty into the inverted Vs profiles \textbf{\textit{considering multiple parameterizations at a real site}}. The comparison is made on the basis of: (a) the residual mean ($\mu_{res}$) [i.e., the difference between the mean of the inverted theoretical dispersion curves ($\mu_{inv}$) and the mean of the measured experimental dispersion data ($\mu_{meas}$)] normalized by $\mu_{meas}$ in percent, (d) the residual coefficient of variation [i.e., the difference between the coefficient of variation of the inverted theoretical dispersion curves ($\delta_{inv}$) and the coefficient of variation of the measured experimental dispersion data ($\delta_{meas}$)]. The vertical dashed lines in panels (a) and (d) at approximately 3, 6, 9, and 14 Hz denote the locations of the ``slices'' shown in panels (b), (c), (e), and (f), respectively. These ``slices'' compare the measured experimental dispersion data's distribution (solid black line) with the distribution of Rayleigh wave velocity (Vr) derived from inversion (histogram).} 
	\label{fig:14}
\end{figure}

The frequency-dependent uncertainty in the experimental dispersion data was propagated through the inversion process using the newly proposed procedure outlined in Figure \ref{fig:6}. To incorporate the additional epistemic uncertainty from the inversion's parameterization, five LN-type parameterizations using LN = 3, 5, 7, 9, and 14 were considered following the same procedure as presented in the previous synthetic example. Inversions took 2 hours to complete using 5 SKX nodes on the Stampede2 cluster. After inversion, the parameterization-quality criteria proposed by Vantassel and Cox (\citeyear{vantassel_swinvert_2021}) was used to assess the performance of the five parameterizations. The criteria showed the LN=3 parameterization underperformed its counterparts and was therefore removed from further consideration to avoid biasing the results. The resulting 1000 theoretical dispersion curves (i.e., 250 curves/parameterization * 4 acceptable parameterizations = 1000 curves) that were fit to the 1000 experimental dispersion data realizations are shown alongside the experimental dispersion data in Figure \ref{fig:13}b. Note that the theoretical curves in Figure13b have been colored in terms of their dispersion misfit relative to the original experimental data, and not their respective realizations. Again, we observe that the majority of the curves have a misfit less than 1.0 (i.e., are on average within 1 standard deviation of the mean), with quite a few below 0.2 that closely follow the mean experimental dispersion trend. Importantly, however, a number of curves with misfits greater than 1.0 are also present, as should be expected if the dispersion uncertainty is accurately being represented in the inversion results. This leads to a qualitative comparison that the experimental dispersion data's uncertainty and the uncertainty implied by the suites of theoretical dispersion curves derived from the inversion are in excellent agreement. Figure \ref{fig:14} presents a quantitative comparison between the uncertainty of the experimental dispersion data and inverted theoretical dispersion curves, highlighting four different ``slices'' using vertical colored dashed lines at approximately 3, 6, 13, and 30 Hz. The quantitative comparison confirms that the inversion-derived models are accurately capturing the experimental dispersion uncertainty in terms of their mean (Figure \ref{fig:14}a), standard deviation (Figure \ref{fig:14}d), and distribution shapes (Figures \ref{fig:14}b, c, e, and f). It is important to note that the minor discrepancies between the measured and inverted data's mean value and uncertainty (refer to Figure \ref{fig:14}a and \ref{fig:14}d) are to be expected, because unlike synthetic data, real experimental data may have some error in its mean and is likely to have variable frequency-dependent uncertainty, making the exact replication of some realizations difficult. For example, the dispersion COVs at this site varied between 0.02 -- 0.08, unlike the synthetic examples which had constant dispersion COVs of 0.05.  However, despite these complicating factors the new procedure is able to produce results that fit the experimental dispersion data remarkably well, with good agreement being observed between the measured and inverted dispersion uncertainty across the entire bandwidth of interest.

\begin{figure}[!t]
	\centering
	\includegraphics{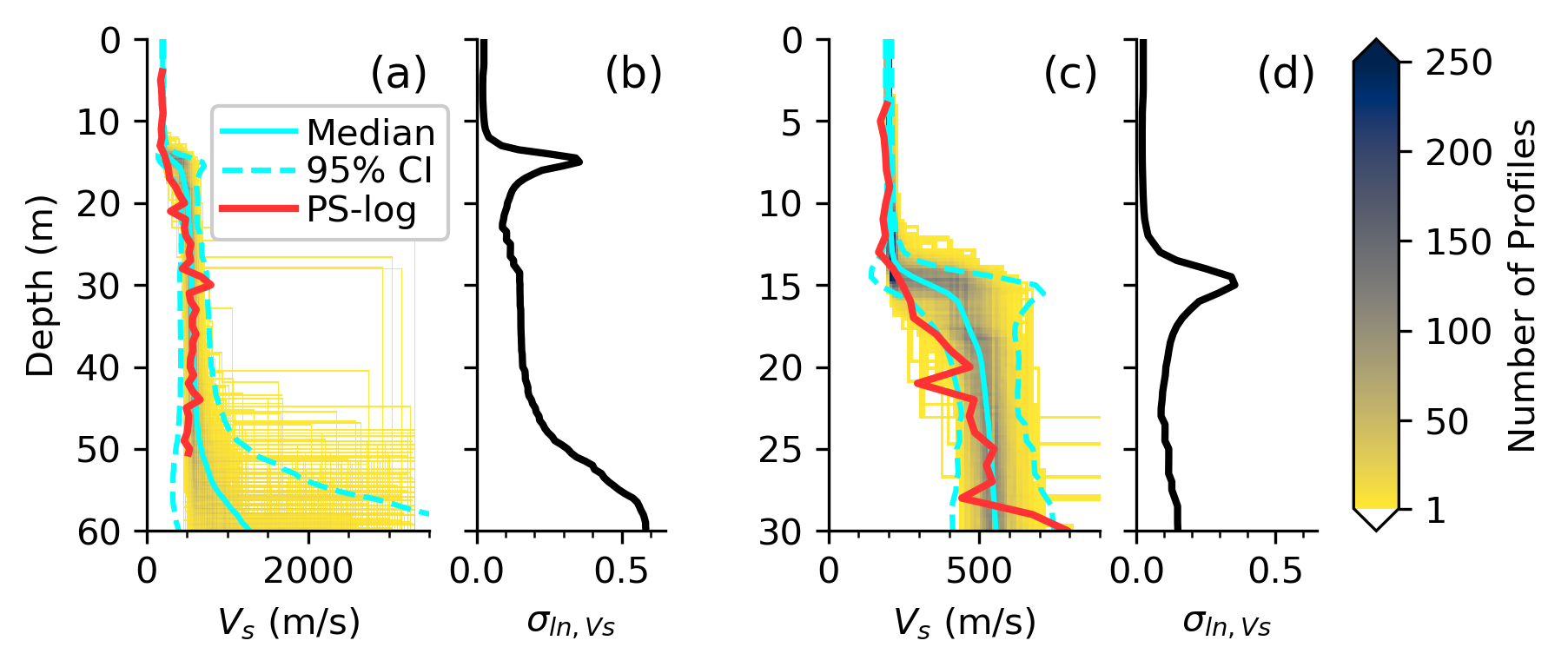}
	\caption{Uncertainty-consistent Vs profiles \textbf{\textit{considering multiple inversion parameterizations at a real site}} using the \textbf{\textit{newly proposed procedure}}. Panels (a) and (c), summarize the Vs profiles resulting from the inversion of the 1000 realized experimental dispersion curves (4 parameterizations with N=250 realizations each) at two separate depth scales. In order to better illustrate their concentration, the profiles have been discretized in terms of depth and Vs, binned into cells, and color mapped in terms of the number of profiles in each cell. The profiles are shown alongside their discretized log-normal median and 95\% confidence interval (CI) and a P-S-suspension log from nearby. Panels (b) and (c), show the lognormal standard deviation of Vs ($\sigma_{ln,Vs}$) of the 1000 profiles shown in panel (a) and (c), respectively. The sharp spike in $\sigma_{ln,Vs}$ at approximately 15 m is due to the uncertainty in the profile's layer boundaries, reflecting a shortcoming in how $\sigma_{ln,Vs}$ has been calculated historically, and is not the result of actual uncertainty in Vs directly.}
	\label{fig:15}
\end{figure}

Figure \ref{fig:15} shows the inversion-derived, uncertainty-consistent Vs profiles at two depth scales; Figure \ref{fig:15}a to a depth of 60 m and Figure \ref{fig:15}b to a depth of 30 m. The Vs profiles reveal a three-layered system consistent with the anticipated geologic conditions and comprised of approximately 15 m of soft soil (Vs of approximately 200 m/s), overlying at least 35 m of weathered rock (Vs of approximately 600 m/s), overlying stiffer material. The uncertainty in the Vs profiles is quantified using $\sigma_{ln,Vs}$ and presented in Figures \ref{fig:15}b and \ref{fig:15}d at the same depth scales as those for Vs (i.e., Figures \ref{fig:15}a and \ref{fig:15}c). $\sigma_{ln,Vs}$ increases from close to zero for the surface layer, where the high frequency/short wavelength surface-wave dispersion data had low COVs, to a $\sigma_{ln,Vs}$ of  0.6 at depth, where due to the lack of very low frequency/long wavelength data coupled with relatively high dispersion COVs the velocity of the material cannot be accurately resolved. We consider the depth where the Vs uncertainty becomes so large as to make meaningful inferences about the site's stiffness intractable as the maximum depth limitation of the presented Vs profiles (i.e., a depth of approximately 45 m, where $\sigma_{ln,Vs}$ starts to exceed about 0.2). To illustrate the accuracy of the surface wave measurements, a P-S-suspension log from the site is shown in comparison to the Vs profiles in Figures 15a and 15c. The P-S-suspension log is seen to reside solely within the 95\% CI of the inverted Vs profiles, despite two complicating factors. First, the inversion-derived Vs profiles are the result of measurements made over a 70 m by 70 m area with known spatial variability, whereas the P-S-suspension log is more-or-less a point measurement. And second, the P-S-suspension log was performed at the western edge of the 70 m by 70 m site rather than at its center due to site constraints. When these two factors are combined, one would expect some differences between the invasive and non-invasive Vs profiles. Nonetheless, the agreement is quite good and the Vs uncertainty associated with the surface wave profiles has been robustly quantified for use in subsequent engineering analyses, while that of the single P-S log is unknown and would have to be assumed. This case study illustrates that the proposed procedure can rigorously propagate experimental dispersion uncertainty through the surface wave inversion process and produce suites of Vs profiles that, while uncertain, can produce results that are consistent with invasive characterization methods.

\section{Conclusion}

Surface wave methods are increasingly being preferred over traditional site characterization methods for measuring a site's shear-stiffness due to their non-invasive nature. The quantification and propagation of uncertainty from non-invasive measurements into Vs profiles for use in subsequent engineering analyses has been the focus of much work in recent years, and while significant progress has been made, no study has yet been able to show quantitative evidence of the successful propagation of experimental dispersion uncertainty into the resulting Vs profiles. Therefore, this study began by examining methods currently available from the literature and showing their deficiencies in three specific categories. First, current methods are shown to be highly sensitive to their many user-defined inversion input parameters (e.g., number of global-inversion trial models), making it difficult/impossible for them to be performed repeatedly by different analysts. Second, the suites of inverted Vs profiles derived from these methods, when viewed in terms of their implied theoretical dispersion data, are shown to systematically under-estimate the uncertainty present in the experimental dispersion data, though they may appear satisfactory when viewed purely qualitatively. And third, if the uncertainties in the implied theoretical dispersion data were to be examined quantitatively, which has not been done previously, there is no obvious remedy available to the analyst to resolve any inconsistency between the measured and inverted dispersion uncertainty. Therefore, a new approach has been proposed that seeks to remedy these shortcomings. First, in addition to the necessary considerations for any inversion, the method is governed by only one additional user-defined input parameter (N), which controls the number of dispersion curve realizations. A value of N=250 has been shown to perform well throughout this study, however, as N is expected to be problem dependent the selection of an appropriate value of N is left to the analyst. Second, the new method requires the quantitative comparison between the measured and inverted dispersion uncertainties to ensure suites of Vs profiles which quantitatively reproduce the uncertainties in the experimental dispersion data. And third, should the measured and inverted dispersion uncertainty be in disagreement, specific guidance is provided on the actions necessary (e.g., increasing the value of N) to produce Vs profiles that better account for the experimental uncertainty. The application of the new procedure has been demonstrated using two synthetic tests and a real-world example. Results show the procedure's ability to produce suites of Vs profiles which accurately capture the site's Vs structure, while rigorously propagating the measured, site-specific dispersion data's uncertainty through the inversion process, yielding uncertainty-consistent Vs profiles that can be used in subsequent engineering analyses.

\section{Acknowledgements}

The first author would like to thank Professor Bob Gilbert for his course Decision, Risk, and Reliability taught at The University of Texas at Austin which inspired the basis of this study. The second author would like to thank numerous colleagues, like Brendon Bradley, Cecile Cornou, Sebastiano Foti, Albert Kottke, Tony Martin, Ellen Rathje, Adrian Rodriguez-Marek, Marc Wathelet, and Liam Wotherspoon and former graduate students Trent Ellis, Shawn Griffiths, David Teague, and Clint Wood who have exchanged thoughts on this topic over the years. All of the inversion results presented in this paper were produced using the Dinver module of the open-source software Geopsy \citep{wathelet_geopsy_2020} which is freely available for download at www.geopsy.org. All inversion pre- and post-processing was performed with the assistance of the open-source Python package SWprepost \citep{vantassel_jpvantasselswprepost_2020}.  Inversions were performed using the Texas Advanced Computing Center's (TACCs) cluster Stampede2 using an allocation provided through DesignSafe-CI \citep{rathje_designsafe_2017}. The figures in this paper were created using Matplotlib 3.1.2 \citep{hunter_matplotlib_2007} and Inkscape 0.92.4.

\bibliographystyle{plainnat}
\bibliography{vs_uncertainty}

\begin{thebibliography}{39}
\providecommand{\natexlab}[1]{#1}
\providecommand{\url}[1]{\texttt{#1}}
\expandafter\ifx\csname urlstyle\endcsname\relax
  \providecommand{\doi}[1]{doi: #1}\else
  \providecommand{\doi}{doi: \begingroup \urlstyle{rm}\Url}\fi

\bibitem[Andrus and Stokoe(2000)]{andrus_liquefaction_2000}
Ronald~D. Andrus and Kenneth~H. Stokoe.
\newblock Liquefaction {Resistance} of {Soils} from {Shear}-{Wave} {Velocity}.
\newblock \emph{Journal of Geotechnical and Geoenvironmental Engineering},
  126\penalty0 (11):\penalty0 1015--1025, November 2000.
\newblock ISSN 1090-0241, 1943-5606.
\newblock \doi{10.1061/(ASCE)1090-0241(2000)126:11(1015)}.

\bibitem[Cox and Wood(2011)]{cox_surface_2011}
B.~R. Cox and C.~M. Wood.
\newblock Surface {Wave} {Benchmarking} {Exercise}: {Methodologies}, {Results},
  and {Uncertainties}.
\newblock In \emph{{GeoRisk} 2011}, pages 845--852, Atlanta, Georgia, United
  States, June 2011. American Society of Civil Engineers.
\newblock ISBN 978-0-7844-1183-4.
\newblock \doi{10.1061/41183(418)89}.

\bibitem[Cox et~al.(2014)Cox, Wood, and Teague]{cox_synthesis_2014}
B.~R. Cox, C.~M. Wood, and David~P. Teague.
\newblock Synthesis of the {UTexas1} {Surface} {Wave} {Dataset}
  {Blind}-{Analysis} {Study}: {Inter}-{Analyst} {Dispersion} and {Shear} {Wave}
  {Velocity} {Uncertainty}.
\newblock In \emph{Geo-{Congress} 2014 {Technical} {Papers}}, pages 850--859,
  Atlanta, Georgia, February 2014. American Society of Civil Engineers.
\newblock ISBN 978-0-7844-1327-2.
\newblock \doi{10.1061/9780784413272.083}.

\bibitem[Cox and Vantassel(2018)]{cox_dynamic_2018}
Brady Cox and Joseph Vantassel.
\newblock Dynamic {Characterization} of {Wellington}, {New} {Zealand}.
\newblock \emph{DesignSafe-CI [publisher]}, November 2018.
\newblock \doi{10.17603/DS24M6J}.
\newblock Type: dataset.

\bibitem[Cox and Teague(2016)]{cox_layering_2016}
Brady~R. Cox and David~P. Teague.
\newblock Layering ratios: a systematic approach to the inversion of surface
  wave data in the absence of a priori information.
\newblock \emph{Geophysical Journal International}, 207\penalty0 (1):\penalty0
  422--438, October 2016.
\newblock ISSN 0956-540X, 1365-246X.
\newblock \doi{10.1093/gji/ggw282}.

\bibitem[Deschenes et~al.(2018)Deschenes, Wood, Wotherspoon, Bradley, and
  Thomson]{deschenes_development_2018}
Michael~R. Deschenes, Clinton~M. Wood, Liam~M. Wotherspoon, Brendon~A. Bradley,
  and Ethan Thomson.
\newblock Development of {Deep} {Shear} {Wave} {Velocity} {Profiles} in the
  {Canterbury} {Plains}, {New} {Zealand}.
\newblock \emph{Earthquake Spectra}, 34\penalty0 (3):\penalty0 1065--1089,
  August 2018.
\newblock ISSN 8755-2930.
\newblock \doi{10.1193/122717EQS267M}.

\bibitem[Di~Giulio et~al.(2012)Di~Giulio, Savvaidis, Ohrnberger, Wathelet,
  Cornou, Knapmeyer-Endrun, Renalier, Theodoulidis, and
  Bard]{di_giulio_exploring_2012}
Giuseppe Di~Giulio, Alexandros Savvaidis, Matthias Ohrnberger, Marc Wathelet,
  Cecile Cornou, Brigitte Knapmeyer-Endrun, Florence Renalier, Nikos
  Theodoulidis, and Pierre-Yves Bard.
\newblock Exploring the model space and ranking a best class of models in
  surface-wave dispersion inversion: {Application} at {European} strong-motion
  sites.
\newblock \emph{Geophysics}, 77\penalty0 (3):\penalty0 B147--B166, May 2012.
\newblock ISSN 0016-8033, 1942-2156.
\newblock \doi{10.1190/geo2011-0116.1}.

\bibitem[EPRI(2012)]{epri_seismic_2012}
EPRI.
\newblock Seismic {Evaluation} {Guidance}: {Screening}, {Prioritization} and
  {Implementation} {Details} ({SPID}) for the {Resolution} of {Fukushima}
  {Near}-{Term} {Task} {Force} {Recommendation} 2.1: {Seismic}.
\newblock Technical Report 1025287, Palo Alto, CA, November 2012.

\bibitem[Foti et~al.(2009)Foti, Comina, Boiero, and
  Socco]{foti_non-uniqueness_2009}
S.~Foti, C.~Comina, D.~Boiero, and L.V. Socco.
\newblock Non-uniqueness in surface-wave inversion and consequences on seismic
  site response analyses.
\newblock \emph{Soil Dynamics and Earthquake Engineering}, 29\penalty0
  (6):\penalty0 982--993, June 2009.
\newblock ISSN 02677261.
\newblock \doi{10.1016/j.soildyn.2008.11.004}.

\bibitem[Foti et~al.(2015)Foti, Lai, Rix, and Strobbia]{foti_surface_2015}
Sebastiano Foti, Carlo~G. Lai, Glenn~J. Rix, and Claudio Strobbia.
\newblock \emph{Surface {Wave} {Methods} for {Near}-{Surface} {Site}
  {Charachterization}}.
\newblock CRC Press, Boca Raton, FL, 1 edition, 2015.

\bibitem[Foti et~al.(2018)Foti, Hollender, Garofalo, Albarello, Asten, Bard,
  Comina, Cornou, Cox, Di~Giulio, Forbriger, Hayashi, Lunedei, Martin,
  Mercerat, Ohrnberger, Poggi, Renalier, Sicilia, and
  Socco]{foti_guidelines_2018}
Sebastiano Foti, Fabrice Hollender, Flora Garofalo, Dario Albarello, Michael
  Asten, Pierre-Yves Bard, Cesare Comina, Cécile Cornou, Brady Cox, Giuseppe
  Di~Giulio, Thomas Forbriger, Koichi Hayashi, Enrico Lunedei, Antony Martin,
  Diego Mercerat, Matthias Ohrnberger, Valerio Poggi, Florence Renalier,
  Deborah Sicilia, and Valentina Socco.
\newblock Guidelines for the good practice of surface wave analysis: a product
  of the {InterPACIFIC} project.
\newblock \emph{Bulletin of Earthquake Engineering}, 16\penalty0 (6):\penalty0
  2367--2420, June 2018.
\newblock ISSN 1570-761X, 1573-1456.
\newblock \doi{10.1007/s10518-017-0206-7}.

\bibitem[Garofalo et~al.(2016)Garofalo, Foti, Hollender, Bard, Cornou, Cox,
  Dechamp, Ohrnberger, Perron, Sicilia, Teague, and
  Vergniault]{garofalo_interpacific_2016}
F.~Garofalo, S.~Foti, F.~Hollender, P.Y. Bard, C.~Cornou, B.R. Cox, A.~Dechamp,
  M.~Ohrnberger, V.~Perron, D.~Sicilia, David~P. Teague, and C.~Vergniault.
\newblock {InterPACIFIC} project: {Comparison} of invasive and non-invasive
  methods for seismic site characterization. {Part} {II}: {Inter}-comparison
  between surface-wave and borehole methods.
\newblock \emph{Soil Dynamics and Earthquake Engineering}, 82:\penalty0
  241--254, March 2016.
\newblock ISSN 02677261.
\newblock \doi{10.1016/j.soildyn.2015.12.009}.

\bibitem[Griffiths et~al.(2016{\natexlab{a}})Griffiths, Cox, Rathje, and
  Teague]{griffiths_mapping_2016}
Shawn~C. Griffiths, Brady~R. Cox, Ellen~M. Rathje, and David~P. Teague.
\newblock Mapping {Dispersion} {Misfit} and {Uncertainty} in {Vs} {Profiles} to
  {Variability} in {Site} {Response} {Estimates}.
\newblock \emph{Journal of Geotechnical and Geoenvironmental Engineering},
  142\penalty0 (11):\penalty0 04016062, November 2016{\natexlab{a}}.
\newblock ISSN 1090-0241, 1943-5606.
\newblock \doi{10.1061/(ASCE)GT.1943-5606.0001553}.

\bibitem[Griffiths et~al.(2016{\natexlab{b}})Griffiths, Cox, Rathje, and
  Teague]{griffiths_surface-wave_2016}
Shawn~C. Griffiths, Brady~R. Cox, Ellen~M. Rathje, and David~P. Teague.
\newblock Surface-{Wave} {Dispersion} {Approach} for {Evaluating} {Statistical}
  {Models} {That} {Account} for {Shear}-{Wave} {Velocity} {Uncertainty}.
\newblock \emph{Journal of Geotechnical and Geoenvironmental Engineering},
  142\penalty0 (11):\penalty0 04016061, November 2016{\natexlab{b}}.
\newblock ISSN 1090-0241, 1943-5606.
\newblock \doi{10.1061/(ASCE)GT.1943-5606.0001552}.

\bibitem[Hollender et~al.(2018)Hollender, Cornou, Dechamp, Oghalaei, Renalier,
  Maufroy, Burnouf, Thomassin, Wathelet, Bard, Boutin, Desbordes,
  Douste-Bacqué, Foundotos, Guyonnet-Benaize, Perron, Régnier, Roullé,
  Langlais, and Sicilia]{hollender_characterization_2018}
Fabrice Hollender, Cécile Cornou, Aline Dechamp, Kaveh Oghalaei, Florence
  Renalier, Emeline Maufroy, Clément Burnouf, Sylvette Thomassin, Marc
  Wathelet, Pierre-Yves Bard, Vincent Boutin, Clément Desbordes, Isabelle
  Douste-Bacqué, Laetitia Foundotos, Cédric Guyonnet-Benaize, Vincent Perron,
  Julie Régnier, Agathe Roullé, Mickael Langlais, and Deborah Sicilia.
\newblock Characterization of site conditions (soil class, {VS30}, velocity
  profiles) for 33 stations from the {French} permanent accelerometric network
  ({RAP}) using surface-wave methods.
\newblock \emph{Bulletin of Earthquake Engineering}, 16\penalty0 (6):\penalty0
  2337--2365, June 2018.
\newblock ISSN 1570-761X, 1573-1456.
\newblock \doi{10.1007/s10518-017-0135-5}.

\bibitem[Hunter(2007)]{hunter_matplotlib_2007}
J.~D. Hunter.
\newblock Matplotlib: {A} {2D} {Graphics} {Environment}.
\newblock \emph{Computing in Science Engineering}, 9\penalty0 (3):\penalty0
  90--95, May 2007.
\newblock ISSN 1558-366X.
\newblock \doi{10.1109/MCSE.2007.55}.

\bibitem[Kayen et~al.(2013)Kayen, Moss, Thompson, Seed, Cetin, Kiureghian,
  Tanaka, and Tokimatsu]{kayen_shear-wave_2013}
R.~Kayen, R.~E.~S. Moss, E.~M. Thompson, R.~B. Seed, K.~O. Cetin, A.~Der
  Kiureghian, Y.~Tanaka, and K.~Tokimatsu.
\newblock Shear-{Wave} {Velocity}–{Based} {Probabilistic} and {Deterministic}
  {Assessment} of {Seismic} {Soil} {Liquefaction} {Potential}.
\newblock \emph{Journal of Geotechnical and Geoenvironmental Engineering},
  139\penalty0 (3):\penalty0 407--419, March 2013.
\newblock ISSN 1090-0241, 1943-5606.
\newblock \doi{10.1061/(ASCE)GT.1943-5606.0000743}.

\bibitem[Lai et~al.(2005)Lai, Foti, and Rix]{lai_propagation_2005}
Carlo~G. Lai, Sebastiano Foti, and Glenn~J. Rix.
\newblock Propagation of {Data} {Uncertainty} in {Surface} {Wave} {Inversion}.
\newblock \emph{Journal of Environmental and Engineering Geophysics},
  10\penalty0 (2):\penalty0 219--228, June 2005.
\newblock ISSN 1083-1363.
\newblock \doi{10.2113/JEEG10.2.219}.

\bibitem[Passeri et~al.(2019)Passeri, Foti, Cox, and
  Rodriguez-Marek]{passeri_influence_2019}
Federico Passeri, Sebastiano Foti, Brady~R. Cox, and Adrian Rodriguez-Marek.
\newblock Influence of {Epistemic} {Uncertainty} in {Shear} {Wave} {Velocity}
  on {Seismic} {Ground} {Response} {Analyses}.
\newblock \emph{Earthquake Spectra}, 35\penalty0 (2):\penalty0 929--954, May
  2019.
\newblock ISSN 8755-2930, 1944-8201.
\newblock \doi{10.1193/011018EQS005M}.

\bibitem[Passeri et~al.(2020)Passeri, Foti, and
  Rodriguez-Marek]{passeri_new_2020}
Federico Passeri, Sebastiano Foti, and Adrian Rodriguez-Marek.
\newblock A new geostatistical model for shear wave velocity profiles.
\newblock \emph{Soil Dynamics and Earthquake Engineering}, 136:\penalty0
  106247, 2020.
\newblock ISSN 0267-7261.
\newblock \doi{https://doi.org/10.1016/j.soildyn.2020.106247}.

\bibitem[Rathje et~al.(2010)Rathje, Kottke, and Trent]{rathje_influence_2010}
Ellen~M. Rathje, Albert~R. Kottke, and Whitney~L. Trent.
\newblock Influence of {Input} {Motion} and {Site} {Property} {Variabilities}
  on {Seismic} {Site} {Response} {Analysis}.
\newblock \emph{Journal of Geotechnical and Geoenvironmental Engineering},
  136\penalty0 (4):\penalty0 607--619, April 2010.
\newblock ISSN 1090-0241, 1943-5606.
\newblock \doi{10.1061/(ASCE)GT.1943-5606.0000255}.

\bibitem[Rathje et~al.(2017)Rathje, Dawson, Padgett, Pinelli, Stanzione, Adair,
  Arduino, Brandenberg, Cockerill, Dey, Esteva, Haan, Hanlon, Kareem, Lowes,
  Mock, and Mosqueda]{rathje_designsafe_2017}
Ellen~M. Rathje, Clint Dawson, Jamie~E. Padgett, Jean-Paul Pinelli, Dan
  Stanzione, Ashley Adair, Pedro Arduino, Scott~J. Brandenberg, Tim Cockerill,
  Charlie Dey, Maria Esteva, Fred~L. Haan, Matthew Hanlon, Ahsan Kareem, Laura
  Lowes, Stephen Mock, and Gilberto Mosqueda.
\newblock {DesignSafe}: {New} {Cyberinfrastructure} for {Natural} {Hazards}
  {Engineering}.
\newblock \emph{Natural Hazards Review}, 18\penalty0 (3):\penalty0 06017001,
  August 2017.
\newblock ISSN 1527-6988, 1527-6996.
\newblock \doi{10.1061/(ASCE)NH.1527-6996.0000246}.

\bibitem[Sambridge(1999)]{sambridge_geophysical_1999}
Malcolm Sambridge.
\newblock Geophysical inversion with a neighbourhood algorithm-{I}. {Searching}
  a parameter space.
\newblock \emph{Geophysical Journal International}, 138\penalty0 (2):\penalty0
  479--494, August 1999.
\newblock ISSN 0956540X, 1365246X.
\newblock \doi{10.1046/j.1365-246X.1999.00876.x}.

\bibitem[Stewart et~al.(2014)Stewart, Afshari, and
  Hashash]{stewart_guidelines_2014}
Jonathan~P Stewart, Kioumars Afshari, and Youssef M~A Hashash.
\newblock Guidelines for {Performing} {Hazard}-{Consistent} {One}-{Dimensional}
  {Ground} {Response} {Analysis} for {Ground} {Motion} {Prediction}.
\newblock Technical Report 2014/16, PEER, Pacific Earthquake Engineering
  Research Center Headquarters, University of California at Berkeley, October
  2014.

\bibitem[Teague and Cox(2016)]{teague_site_2016}
David~P. Teague and Brady~R. Cox.
\newblock Site response implications associated with using non-unique {Vs}
  profiles from surface wave inversion in comparison with other commonly used
  methods of accounting for {Vs} uncertainty.
\newblock \emph{Soil Dynamics and Earthquake Engineering}, 91:\penalty0
  87--103, December 2016.
\newblock ISSN 02677261.
\newblock \doi{10.1016/j.soildyn.2016.07.028}.

\bibitem[Teague et~al.(2018{\natexlab{a}})Teague, Cox, Bradley, and
  Wotherspoon]{teague_development_2018}
David~P. Teague, Brady~R. Cox, Brendon Bradley, and Liam Wotherspoon.
\newblock Development of {Deep} {Shear} {Wave} {Velocity} {Profiles} with
  {Estimates} of {Uncertainty} in the {Complex} {Interbedded} {Geology} of
  {Christchurch}, {New} {Zealand}.
\newblock \emph{Earthquake Spectra}, 34\penalty0 (2):\penalty0 639--672, May
  2018{\natexlab{a}}.
\newblock ISSN 8755-2930.
\newblock \doi{10.1193/041117EQS069M}.

\bibitem[Teague et~al.(2018{\natexlab{b}})Teague, Cox, and
  Rathje]{teague_measured_2018}
David~P. Teague, Brady~R. Cox, and Ellen~M. Rathje.
\newblock Measured vs. predicted site response at the {Garner} {Valley}
  {Downhole} {Array} considering shear wave velocity uncertainty from borehole
  and surface wave methods.
\newblock \emph{Soil Dynamics and Earthquake Engineering}, 113:\penalty0
  339--355, October 2018{\natexlab{b}}.
\newblock ISSN 02677261.
\newblock \doi{10.1016/j.soildyn.2018.05.031}.

\bibitem[Toro(1995)]{toro_probabilistic_1995}
Gabriel Toro, R.
\newblock Probabilistic models of site velocity profiles for generic and
  site-specific ground-motion amplification studies.
\newblock Technical report, November 1995.
\newblock Published as Appendix C of Silva et al 1996.

\bibitem[Vantassel(2020)]{vantassel_jpvantasselswprepost_2020}
Joseph Vantassel.
\newblock jpvantassel/swprepost: latest ({Concept}), May 2020.
\newblock URL \url{http://doi.org/10.5281/zenodo.3839998}.

\bibitem[Vantassel et~al.(2018)Vantassel, Cox, Wotherspoon, and
  Stolte]{vantassel_mapping_2018}
Joseph Vantassel, Brady Cox, Liam Wotherspoon, and Andrew Stolte.
\newblock Mapping {Depth} to {Bedrock}, {Shear} {Stiffness}, and {Fundamental}
  {Site} {Period} at {CentrePort}, {Wellington}, {Using} {Surface}‐{Wave}
  {Methods}: {Implications} for {Local} {Seismic} {Site} {Amplification}.
\newblock \emph{Bulletin of the Seismological Society of America}, 108\penalty0
  (3B):\penalty0 1709--1721, May 2018.
\newblock ISSN 0037-1106.
\newblock \doi{10.1785/0120170287}.

\bibitem[Vantassel and Cox(2020)]{vantassel_surface_2020}
Joseph~P. Vantassel and Brady~R. Cox.
\newblock Surface {Wave} {Inversion} {Benchmarks}.
\newblock \emph{DesignSafe-CI [publisher]}, May 2020.
\newblock \doi{10.17603/ds2-cpmr-v194}.
\newblock Type: dataset.

\bibitem[Vantassel and Cox(2021)]{vantassel_swinvert_2021}
Joseph~P. Vantassel and Brady~R. Cox.
\newblock {SWinvert}: a workflow for performing rigorous 1-{D} surface wave
  inversions.
\newblock \emph{Geophysical Journal International}, 224\penalty0 (2):\penalty0
  1141--1156, 2021.
\newblock \doi{10.1093/gji/ggaa426}.
\newblock URL \url{https://academic.oup.com/gji/article/224/2/1141/5903275}.
\newblock Publisher: Oxford University Press.

\bibitem[Wathelet(2008)]{wathelet_improved_2008}
M.~Wathelet.
\newblock An improved neighborhood algorithm: {Parameter} conditions and
  dynamic scaling.
\newblock \emph{Geophysical Research Letters}, 35\penalty0 (9):\penalty0
  L09301, May 2008.
\newblock ISSN 0094-8276.
\newblock \doi{10.1029/2008GL033256}.

\bibitem[Wathelet et~al.(2004)Wathelet, Jongmans, and
  Ohrnberger]{wathelet_surface-wave_2004}
M.~Wathelet, D.~Jongmans, and M.~Ohrnberger.
\newblock Surface-wave inversion using a direct search algorithm and its
  application to ambient vibration measurements.
\newblock \emph{Near Surface Geophysics}, 2\penalty0 (4):\penalty0 211--221,
  November 2004.
\newblock ISSN 15694445.
\newblock \doi{10.3997/1873-0604.2004018}.

\bibitem[Wathelet et~al.(2018)Wathelet, Guillier, Roux, Cornou, and
  Ohrnberger]{wathelet_rayleigh_2018}
M~Wathelet, B~Guillier, P~Roux, C~Cornou, and M~Ohrnberger.
\newblock Rayleigh wave three-component beamforming: signed ellipticity
  assessment from high-resolution frequency-wavenumber processing of ambient
  vibration arrays.
\newblock \emph{Geophysical Journal International}, 215\penalty0 (1):\penalty0
  507--523, October 2018.
\newblock ISSN 0956-540X, 1365-246X.
\newblock \doi{10.1093/gji/ggy286}.

\bibitem[Wathelet(2005)]{wathelet_array_2005}
Marc Wathelet.
\newblock \emph{Array recordings of ambient vibrations: surface wave
  inversion}.
\newblock PhD thesis, Universite de Liege Faculte des Sciences Aplliquees,
  Liege, Belgium, February 2005.

\bibitem[Wathelet et~al.(2020)Wathelet, Chatelain, Cornou, Giulio, Guillier,
  Ohrnberger, and Savvaidis]{wathelet_geopsy_2020}
Marc Wathelet, Jean-Luc Chatelain, Cécile Cornou, Giuseppe~Di Giulio, Bertrand
  Guillier, Matthias Ohrnberger, and Alexandros Savvaidis.
\newblock Geopsy: {A} {User}-{Friendly} {Open}-{Source} {Tool} {Set} for
  {Ambient} {Vibration} {Processing}.
\newblock \emph{Seismological Research Letters}, April 2020.
\newblock ISSN 0895-0695, 1938-2057.
\newblock \doi{10.1785/0220190360}.

\bibitem[Wood et~al.(2017)Wood, Cox, Green, Wotherspoon, Bradley, and
  Cubrinovski]{wood_vs-based_2017}
Clinton~M. Wood, Brady~R. Cox, Russell~A. Green, Liam~M. Wotherspoon,
  Brendon~A. Bradley, and Misko Cubrinovski.
\newblock Vs-{Based} {Evaluation} of {Select} {Liquefaction} {Case} {Histories}
  from the 2010–2011 {Canterbury} {Earthquake} {Sequence}.
\newblock \emph{Journal of Geotechnical and Geoenvironmental Engineering},
  143\penalty0 (9):\penalty0 04017066, September 2017.
\newblock ISSN 1090-0241, 1943-5606.
\newblock \doi{10.1061/(ASCE)GT.1943-5606.0001754}.

\bibitem[Yust et~al.(2018)Yust, Cox, and Cheng]{yust_epistemic_2018}
Michael Yust, Brady~R. Cox, and Tianjian Cheng.
\newblock Epistemic {Uncertainty} in {Vs} {Profiles} and {Vs30} {Values}
  {Derived} from {Joint} {Consideration} of {Surface} {Wave} and {H}/{V} {Data}
  at the {FW07} {TexNet} {Station}.
\newblock pages 387--399, Austin, Texas, June 2018.
\newblock \doi{https://doi.org/10.1061/9780784481462.038}.

\end{thebibliography}

\end{document}